\documentclass[aps,prd,nofootinbib,twocolumn,superscriptaddress,preprintnumbers,balancelastpage,longbibliography]{revtex4-1}

\usepackage{amsmath,amssymb,mathtools,bm}
\usepackage{graphicx, color, hepunits}
\usepackage[dvipsnames]{xcolor}
\usepackage{float}
\usepackage{multirow}
 \usepackage{hyperref}
\hypersetup{
    colorlinks=true,       
    linkcolor=blue,        
    citecolor=blue,        
    filecolor=magenta,     
    urlcolor=blue          
}
\usepackage[utf8]{inputenc}
\usepackage[english]{babel}
\usepackage[nolist]{acronym}
\usepackage{xfrac}
\usepackage{xspace}
\usepackage{tabularx}
\usepackage{array}
\usepackage{makecell}

\usepackage{listings}

\newcommand{\es}[2] {\begin{equation} \label{#1} \begin{split} #2 \end{split} \end{equation}}

\newcolumntype{C}[1]{>{\centering\let\newline\\\arraybackslash\hspace{0pt}}m{#1}}
\newcolumntype{P}[1]{>{\centering\arraybackslash}p{#1}}
\newcommand{\ra}[1]{\renewcommand{\arraystretch}{#1}}

\newcommand{\amrex}{AMReX}

\begin{document}

\title{
Dark Matter from Axion Strings with Adaptive Mesh Refinement
}
\date{\today}
\author{Malte Buschmann}
\email{msab@princeton.edu}
\affiliation{Department of Physics, Princeton University, Princeton, NJ 08544, USA}

\author{Joshua~W.~Foster}
\email{jwfoster@mit.edu}
\affiliation{Leinweber Center for Theoretical Physics, Department of Physics, University of Michigan, Ann Arbor, MI 48109}

\affiliation{Berkeley Center for Theoretical Physics, University of California, Berkeley, CA 94720}
\affiliation{Theoretical Physics Group, Lawrence Berkeley National Laboratory, Berkeley, CA 94720}

\author{Anson Hook}
\affiliation{Maryland Center for Fundamental Physics, University of Maryland, College Park, MD 20742, U.S.A.}

\author{Adam Peterson}
\affiliation{Center for Computational Sciences and Engineering
Lawrence Berkeley National Laboratory
Berkeley, CA 94720}

\author{Don E. Willcox}
\affiliation{Center for Computational Sciences and Engineering
Lawrence Berkeley National Laboratory
Berkeley, CA 94720}

\author{Weiqun Zhang}
\affiliation{Center for Computational Sciences and Engineering
Lawrence Berkeley National Laboratory
Berkeley, CA 94720}

\author{Benjamin~R.~Safdi}
\email{brsafdi@berkeley.edu}
\affiliation{Berkeley Center for Theoretical Physics, University of California, Berkeley, CA 94720}
\affiliation{Theoretical Physics Group, Lawrence Berkeley National Laboratory, Berkeley, CA 94720}

\begin{abstract}

Axions are hypothetical particles that may explain the observed dark matter (DM) density
and 
the non-observation of a neutron electric dipole moment.
An increasing number of axion laboratory searches are underway worldwide,
but these efforts are made difficult by the fact that the axion mass is largely unconstrained.  If the axion is generated after inflation there is a unique mass that gives rise to the observed DM abundance; due to nonlinearities and topological defects known as strings, computing this mass accurately has been a challenge for four decades.
Recent works, making use of large static lattice simulations, have led to largely disparate predictions for the axion mass, spanning the range from 25 microelectronvolts to over 500 microelectronvolts.  In this work we show that adaptive mesh refinement (AMR) simulations are better suited for axion cosmology than the previously-used static lattice simulations because only the string cores require high spatial resolution. Using dedicated AMR simulations we obtain an over three order of magnitude leap in dynamic range 
and provide evidence that axion strings radiate their energy with a scale-invariant spectrum, to within $\sim$5\% precision, leading to a mass prediction in the range (40,180) microelectronvolts.

\end{abstract}

\maketitle

An outstanding mystery of the Standard Model of particle physics is that the neutron electric dipole moment, which would cause the neutron to precess in the presence of an electric field, appears to be over ten billion times smaller than expected~\cite{nEDM:2020crw}.   Axions were originally invoked as a dynamical solution to this problem; they would 
interact with quantum chromodynamics (QCD) inside of the neutron
so as to remove the electric dipole moment~\cite{Peccei:1977hh,Peccei:1977ur,Weinberg:1977ma,Wilczek:1977pj}.  However, free-streaming ultra-cold axions may also be produced cosmologically  in the early Universe, and these axions may explain the observed dark matter (DM)~\cite{Preskill:1982cy,Abbott:1982af,Dine:1982ah}, which is known to govern the dynamics of galaxies and galaxy clusters.  

Multiple efforts are underway at present to search for the existence of axion DM in the laboratory~\cite{Graham:2015ouw,Sikivie:2020zpn}, but these efforts are hindered by the fact that the mass of the axion particle is currently unknown.  The axion is naturally realized as the pseudo-Goldstone boson of a global symmetry called the Peccei-Quinn (PQ) symmetry, which is broken at a high energy scale $f_a$~\cite{Peccei:1977hh,Peccei:1977ur,Weinberg:1977ma,Wilczek:1977pj,DiLuzio:2020wdo}.  If the PQ symmetry is broken after the cosmological epoch of inflation, then there is a unique axion mass $m_a$ that leads to the observed DM abundance. (If the PQ symmetry is broken before or during inflation, then the DM abundance depends on the initial value of the axion field that is inflated~\cite{Marsh:2015xka}.)  However, computing this mass is difficult principally because after PQ symmetry breaking axion strings develop; at the string cores the full PQ symmetry is restored.  As the Universe expands the strings shrink, straighten, and combine by emitting radiation into axions.  The contribution to the DM abundance from the string-induced axions has been heavily debated, 
with some works claiming that string-induced axions play no important role~\cite{Harari:1987ht,Hagmann:1990mj}, with the DM abundance dominated by axions produced during the QCD phase transition, and others claiming these axions dominate the DM abundance~\cite{Davis:1989nj,Battye:1993jv,Battye:1994au}.

\begin{figure*}[t]\centering{
\includegraphics[width =.99 \textwidth]{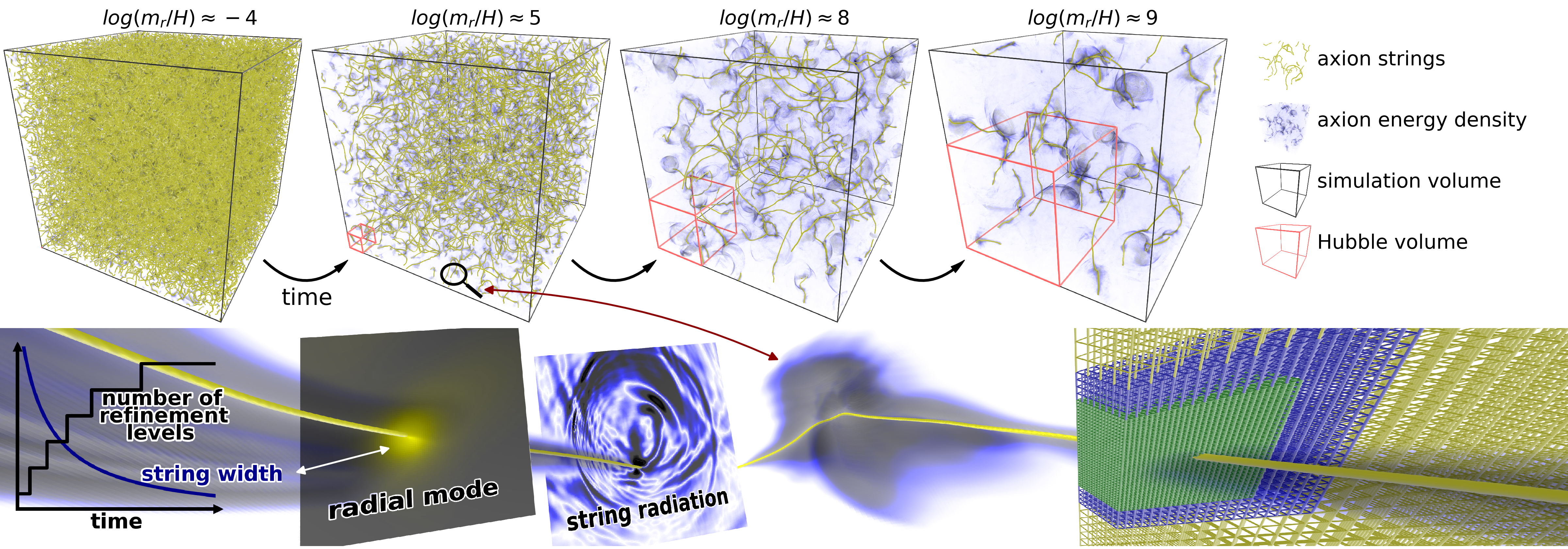}}
\caption{{\sl(Top row)} 3-D rendering of various simulation states from the initial state {\sl(left)} to the final state {\sl(right)}. Shown is the full simulation volume with the respective relative size of a Hubble volume indicated. The axion energy density is illustrated by the density of a 3-D media and string cores are overlaid in yellow. {\sl(Bottom row)} Zoom in on a string segment. {\sl From left to right:} Relationship between the string width and the number of refinement levels as a function of time; 2-D slices of the radial mode and string radiation centered around a string element; string element enshrouded by axion energy density; and an illustration of the layout of the three coarsest grid levels around a string core (not to scale).  Animations available \href{https://bit.ly/amr_axion}{here}.}
\label{fig:Fig1}
\end{figure*}

The evolution of the axion string network in the early Universe has been studied numerically and analytically since the
1980's~\cite{Vilenkin:1982ks,Sikivie:1982qv,Davis:1986xc,Harari:1987ht,Shellard:1987bv,Davis:1989nj,Hagmann:1990mj,Battye:1993jv,Battye:1994au,Yamaguchi:1998gx} with increasingly complex and capable frameworks in recent years~\cite{Klaer:2017ond,Gorghetto:2018myk,Vaquero:2018tib,Buschmann:2019icd,Gorghetto:2020qws,Dine:2020pds}.  The earliest simulations were restricted computationally to lattices of order $\sim$$150^3$ sites~\cite{Davis:1989nj}, while modern-day static-lattice simulations have achieved $\sim$$8,000^3$ sites~\cite{Vaquero:2018tib}.
The approach we present in this work, using adaptive mesh refinement (AMR) simulations, provides an even larger jump in sensitivity by maintaining high resolution around the string cores and lower resolution elsewhere~\cite{Drew:2019mzc}; to achieve the same resolution as our simulations using a static grid would require a $65,536^3$ site lattice.  Our unprecedented dynamical range allows us to determine that radiation from axion strings prior to the QCD phase transition likely dominates the DM density.

\section*{AMR Simulation Framework}

The axion $a$ as the phase of the complex PQ scalar field $\Phi = (r + f_a)/\sqrt{2} e^{i a / f_a}$, with $a = a(x)$ and $r = r(x)$ real functions of spacetime $x$.  The radial mode $r$ is heavy and is not dynamical at temperatures below its mass $m_r$.  The axion field, on the other hand, is massless until the QCD phase transition and thus is dynamical on scales smaller than the cosmological horizon between the PQ and QCD epochs.  The axion field acquires a small mass $m_a \sim \Lambda_{\rm QCD}^2 / f_a$ at temperatures $T$ of order the QCD confinement scale $\Lambda_{\rm QCD}$ from QCD instantons~\cite{diCortona:2015ldu}, though in our simulations we focus on temperatures $T \gg \Lambda_{\rm QCD}$ where the mass may be neglected.

Our simulation is based on the block-structured AMR software framework \amrex~\cite{zhang2020amrex}. The equations of motion (EOM) for $\Phi$ can be derived from the Lagrangian~\cite{Hiramatsu:2012gg}
\begin{equation}
\label{eq:Lagrangian}
\mathcal{L}_{PQ} = | \partial \Phi | ^2 - \lambda \left(|\Phi|^2 - \frac{f_a^2}{2}\right)^2 - \frac{\lambda T^2}{3} |\Phi|^2,
\end{equation} where $\lambda$ is the PQ quartic coupling.  (We fix $\lambda=1$ without loss of generality so that $m_r = \sqrt{2} f_a$.) 
The EOM are solved using the strong-stability preserving Runge-Kutta (SSPRK3) algorithm with a time step size that satisfies the Courant–Friedrichs–Lewy condition  on a lattice defined in fixed comoving coordinates. Evolution takes place in rescaled conformal time $\eta = R/R_1 = (t/t_1)^{1/2}$, where $R$ is the scale factor of the Friedmann–Lemaître–Robertson–Walker metric, $R_1\equiv R(t_1)$, and $t_1$ is a reference time such that $H_1 \equiv H(t_1) = f_a$ with Hubble parameter $H$. In these units the PQ phase transition takes place around $\eta \approx 1$, and we chose a starting time of $\eta_i=0.1$ and a final time of $\eta_f=75.7$. Our simulation volume is a box with periodic boundary conditions and comoving side length $L=120 /(R_1 H_1)$. This volume corresponds to $1200^3$ Hubble volumes at $\eta_i$ and $\sim$$4$ Hubble volumes at $\eta_f$. (Ref.~\cite{Gorghetto:2020qws} found that finite-volume effects are not important for simulations ending with $\sim$4 Hubble volumes.)

The string width scale $\Gamma$ is set by $m_r^{-1}$, while the maximum physical length scale that may be resolved with the comoving lattice grows linearly with $\eta$. 
Thus, finer grids are needed to resolve $\Gamma$  at later times.
We start with a uniform grid of $2048^3$ grid sites, with an initial state based on a thermal distribution before the PQ phase transition (see Methods).
Extra refined grids are then added over time whenever the comoving string width drops below a certain threshold. 
We add the first four extra refinement levels when $\Gamma$ is resolved by four grid sites at the respective finest level, with the fifth extra level added when $\Gamma$ is resolved by three grid sites (see Fig.~\ref{fig:Fig1} and Supp. Fig.~\ref{fig:resolution}). In comparison, note that~\cite{Gorghetto:2020qws} resolves $\Gamma$ by one grid site at the end of their simulation.  
Each extra level introduces eight times as many grid cells per volume as the previous level.
Refined levels are localized primarily around strings. This is achieved by identifying grid cells that are pierced by a string core using the algorithm described in~\cite{Fleury:2015aca}. The exact grid layout is periodically adjusted to track strings over time.
See Fig.~\ref{fig:Fig1} for an illustration of the grid layout. 

\section*{String Network Evolution}

The axion string network is thought to evolve and shrink with time by radiating axions so as to obey the scaling solution, where the number of strings per Hubble patch remains order unity as a function of time~\cite{Davis:1986xc}.  The network evolution is illustrated in the top panels of Fig.~\ref{fig:Fig1},
with time slices labeled by $\log(m_r / H ) = \log(2 m_r t)$.
The energy density in axion radiation is overlaid on top of the string network and is strongest in the vicinity of areas of large string curvature.  

\begin{figure}[t]
\centering{
\includegraphics[width =.49 \textwidth]{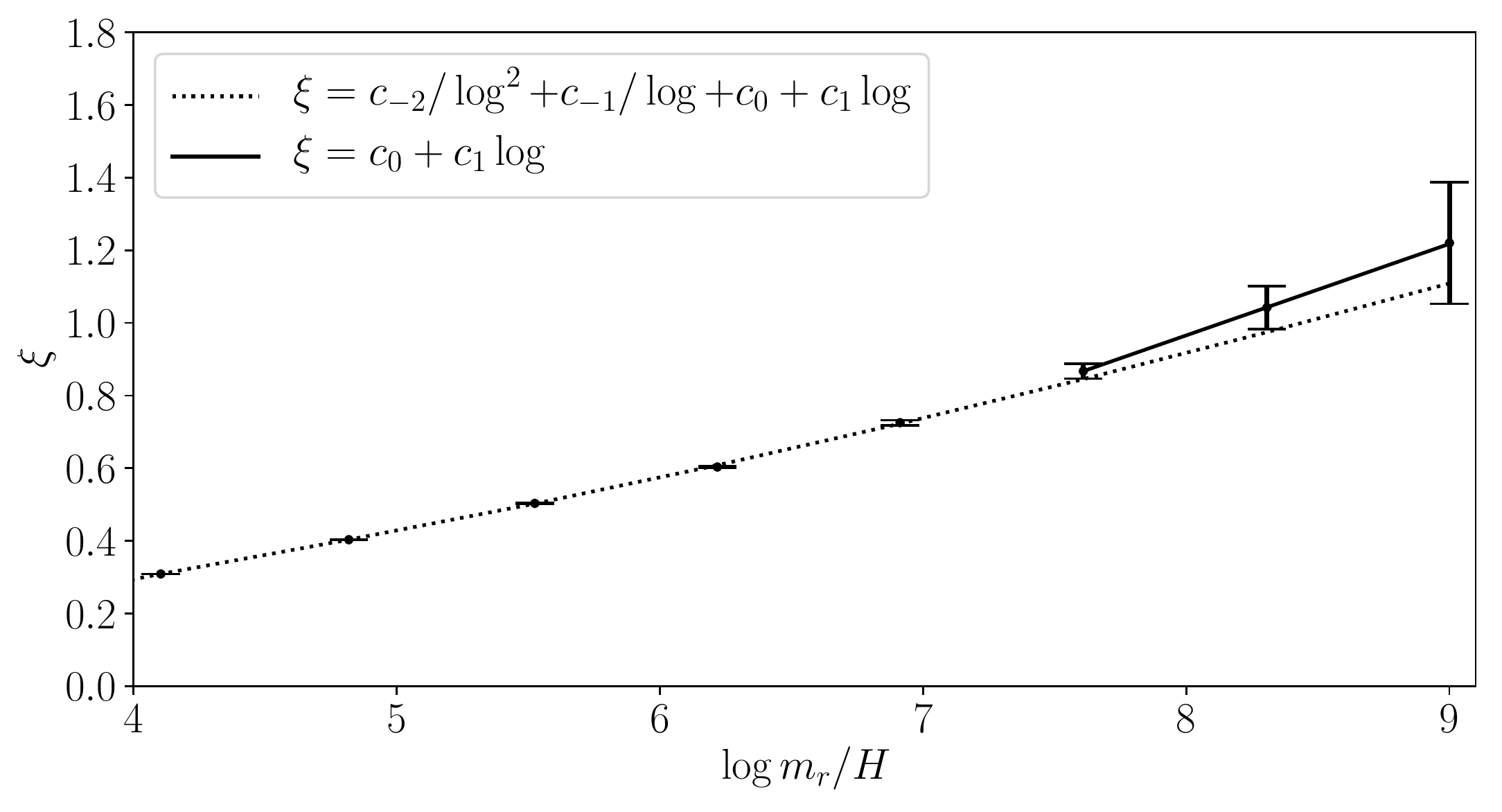}}
\caption{The string length per Hubble volume $\xi$ increases with time in our simulation, indicating a logarithmic violation to the scaling solution~\cite{Gorghetto:2018myk}, which would predict constant $\xi$.
At late times in the simulation (large $\log m_r / H$) the growth in $\xi$ appears linear in $\log m_r / H$ with coefficient $c_1 \approx 0.25$ as measured for the fit over the full $\log m_r / H$ range shown, but including terms all the way down to $c_{-2} / \log^2$. The fit illustrated by the solid curve only includes terms down to $q_0$ but is limited to late times ($\log \in (7.5,9)$); this fit leads to $c_1 \approx 0.25$ also.  These fits indicate that at the beginning of the QCD phase transition, at $\log_* \approx 65$, we expect $\xi_* \approx 15$.    }
\label{fig:xi}
\end{figure}

The string length per Hubble volume is quantified through the parameter  $\xi$, which is defined by $\xi \equiv \ell t^2 / {\mathcal{V}}$ with $\ell$ the total string length in the simulation volume ${\mathcal{V}}$. We determine $\ell$ by counting string-pierced plaquettes in our simulation using the algorithm described in~\cite{Fleury:2015aca}. We illustrate $\xi$ as a function of $\log m_r / H $ in Fig.~\ref{fig:xi}.  
We compute $\xi$ at points in time separated by a Hubble time ($\Delta \log m_r / H = \log 2$), since the network is strongly correlated on time scales smaller than a Hubble time.

We verify that $\xi$ increases linearly with $\log m_r / H$, which was first suggested in~\cite{Gorghetto:2018myk,Gorghetto:2020qws}.
Ref.~\cite{Gorghetto:2020qws} constructed a suite of simulations on static grids of up to $4500^3$ sites and out to at most $\log m_r / H \sim 7.9$; they fit a model of the form $\xi = c_{-2} / \log^2 + c_{-1} / \log + c_0 + c_1 \log$, with $\log \equiv \log m_r / H$, to their $\xi$ data for $\log \in (4.5,7.9)$ and found $c_1 = 0.24\pm 0.02$.  Given $m_r \sim 10^{10}$ GeV and the QCD phase transition beginning at temperatures $T \sim 1$ GeV, the string network is expected to evolve until $\log_*  \sim 65$, which is far beyond the dynamical range that may be simulated. 

In Fig.~\ref{fig:xi} we illustrate our fit of the same functional form as in~\cite{Gorghetto:2020qws} to our $\xi$ data over the range $\log \in (4,9)$; we find $c_0 = -1.82 \pm 0.01$ and $c_1 = 0.254 \pm 0.002$ (see Methods for details). As a systematic test we fit the functional form $\xi =  c_0 + c_1 \log$ to the $\xi$ data over the limited range $\log \in (7.5,9)$ and determine $c_0 \approx -1.05$ and $c_1 \approx 0.252$.  Importantly, the parameter $c_1$, which governs the large $\log$ behavior of $\xi$, agrees between the two methods and agrees with the measurement in~\cite{Gorghetto:2020qws}.  Assuming that the QCD phase transition begins at $\log_* \in (60,70)$ we estimate that at the beginning of the phase transition $\xi = \xi_* \in (13,17)$.
The linear growth of $\xi$ with $\log m_r/ H $ does not support the analytic velocity-dependent one-scale model (see Refs.~\cite{Martins:2018dqg,Hindmarsh:2021vih,Chang:2021afa}), which predicts that $\xi$ should approach a constant at large $\log$.  On the other hand, the observation that $\xi$ grows linearly with $\log$ may be naturally explained by the well-established logarithmic increase of the string tension with time, $\mu(t) \approx \mu_0 \log m_r / H$ with $\mu_0 = \pi f_a^2$ to leading order in large $\log$ (see Supp. Fig.~\ref{fig:string_tension}). A given string segment loses energy at a constant rate that does not evolve with time~\cite{Davis:1986xc}, and as a result energy builds up in the strings relative to the situation where $\mu$ does not increase logarithmically with time.  This increase in energy is manifest by a logarithmically increasing $\xi$. (See Methods for details of this argument.)

\section*{Axion Radiation Spectrum}
 
As the string network evolves in the scaling regime axions are produced at a rate $\Gamma_a \approx 2 H \rho_s$, where $\rho_s = \xi \mu / t^2$ is the energy density in strings.  As we show later in this Article, the DM density from string-induced axion radiation is proportional to the number density of axions at $\log m_r / H = \log_*$.  To compute the number density we need to know the axion radiation spectrum from strings.  We quantify the spectrum through the normalized distribution $F(k/H) = d \log \Gamma_a / d (k/H)$ for physical momentum $k$. (See, {\it e.g.},~\cite{Gorghetto:2018myk} for a review of the analytic aspects of the network evolution.)
We compute $F$ numerically from our simulation ouput by $F(k/H) \propto (1 / R^3) \frac{d}{dt} \big( R^3 \partial\rho_a / \partial k \big)$, with $\partial\rho_a / \partial k$ the time-dependent differential axion energy density spectrum.
 
The axion radiation is distributed in frequency between the effective infrared (IR) cutoff, which is provided by $H$, and the effective ultraviolet (UV) cutoff set by the string width $\sim$$m_r$.  For momenta $k$ well between these two scales ($H \ll k \ll m_r$) the radiation spectrum is expected to follow a power-law.  Below, we describe how we measure the index of this power law.

We calculate $F$ via finite differences in nonuniform $\Delta t$ corresponding to uniform intervals in $\log m_r /H$. In our fiducial analysis, we calculate instantaneous emission spectra using intervals of $\Delta \log m_r /H = 0.25$, which is of order Hubble-time separations. At each $\log m_r / H$ value, we fit a power-law model $F(k/H) \propto 1 / (k/H)^q$ to the instantaneous spectra between an IR cut-off $k_{\rm IR} = x_{\rm IR} H$ and a UV cut-off  $k_{\rm UV} = m_r/x_{\rm UV}$, with the cut-offs chosen to be sufficiently far from the physical IR and UV cut-offs. (See the methods for details of how this fit is performed.)  We chose $x_{\rm IR} = 50$ and $x_{\rm UV} = 16$ in order to be sufficiently far into the power law regime of $k$.

\begin{figure}[htb]
\centering{
\includegraphics[width =.49 \textwidth]{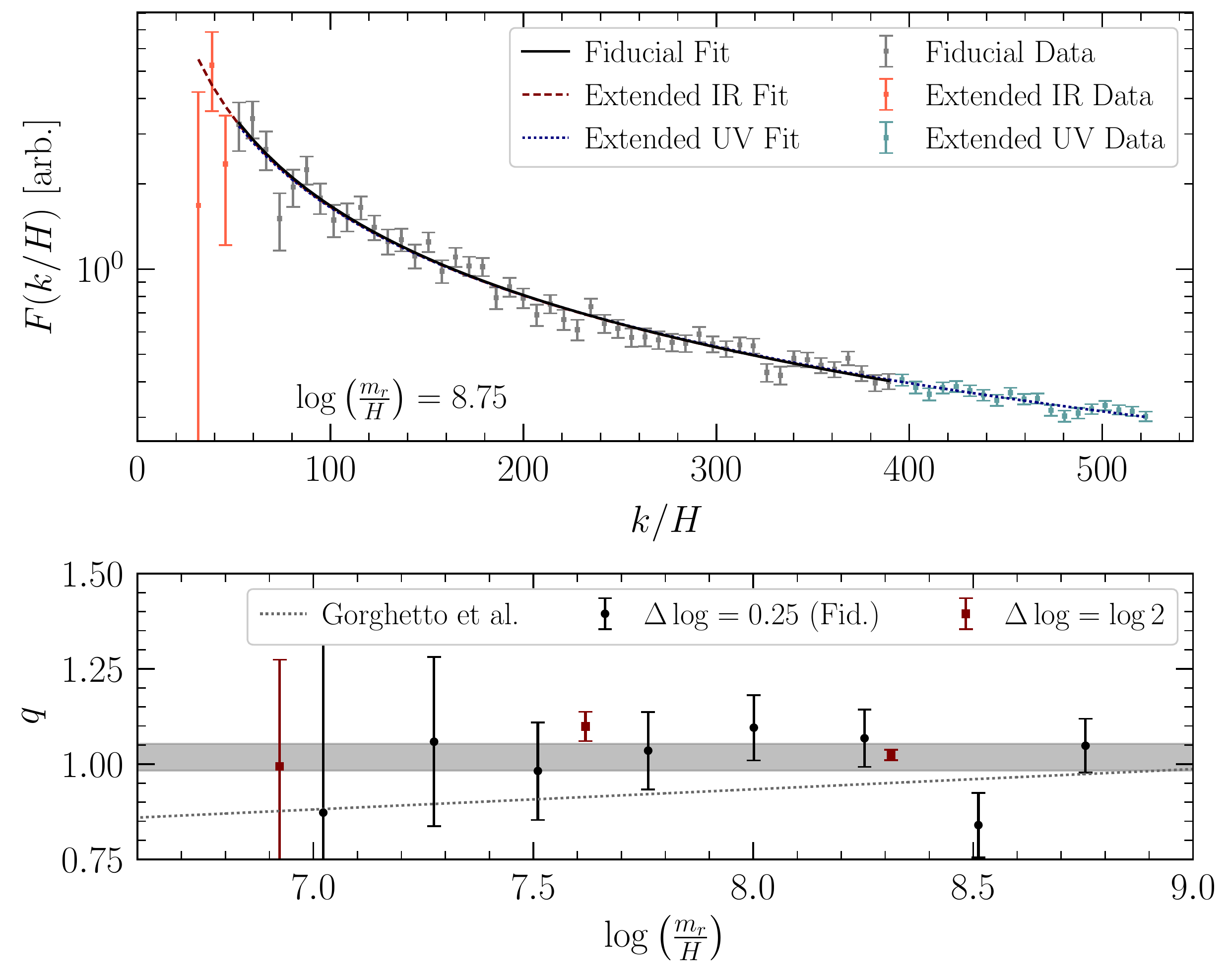}}
\caption{(\textit{Above}) Example fits to the instantaneous emission spectrum calculated at $\log m_r /H  = 8.75$. In our fiducial analysis, the instantaneous emission spectra are calculated using a timestep corresponding to $\Delta \log m_r /H = 0.25$, and a power-law model is fit to the data at $k$ between the IR and UV cutoffs of $k_{\rm IR} = 50H$ and $k_{\rm UV} = m_r / 16$. The data included in this fit range is shown in grey with the best-fit power law depicted in black. We also illustrate two systematic variations, one in which we extend our IR cutoff down to $k_{\rm IR} = 30H$ (``Extended IR Data"), and another where we extend our UV cutoff upward to  $k_{\rm UV} = m_r / 12$ (``Extended UV data").
For clarity, the data are down-binned by a factor of $2$ in $k/H$. (\textit{Below}) The evolution of the fitted power-law index $q$ as a function of $\log m_r /H$. The best fit indices obtained in our fiducial analysis are shown in black, with red showing the indices computed using $\Delta \log m_r /H  = \log 2$.  In our fiducial analysis we constrain $q =1.02 \pm 0.03$, which is shaded. For comparison, the best fit linear growth of $q$ obtained in \cite{Gorghetto:2020qws} is shown in dotted grey.}
\label{fig:SpecIndex}
\end{figure} 

In the top panel of Fig.~\ref{fig:SpecIndex} we illustrate $F$ computed at $\log m_r / H  = 8.75$ for our fiducial choice of $x_{\rm IR}$ and $x_{\rm UV}$ as well as two systematic variations on the choice of fitting range, extending to $x_{\rm IR} = 30$ (``Extended IR Data") and $x_{\rm UV} = 12$ (``Extended UV Data").  The best-fit power-law models are also illustrated. In the bottom panel, we show the evolution of the index $q$ as a function of $\log m_r / H$, for both our fiducial analysis and for a systematic variation where we use $\Delta \log m_r /H = \log 2$ when computing $F$.  We compare our results to the best-fit model obtained in~\cite{Gorghetto:2020qws}, who claimed evidence that $q$ evolves logarithmically in time, with $q > 1$ at late times.  In particular, Ref.~\cite{Gorghetto:2020qws} fit the evolution model $q(t) = q_1 \log(m_r / H) + q_0$ to their $q$ data and found evidence for non-zero $q_1$, claiming $q_1 = 0.053 \pm 0.005$.  Fitting this model to our $q$ data (see Methods for details) yields $q_1 = -0.04 \pm 0.08$ and $q_0 = 1.36 \pm 0.69$, which is in tension with the results in~\cite{Gorghetto:2020qws}. (The best-fit model in that work is inconsistent at the level $\sim$$1.8$$\sigma$ with our measured $q$ values).  Given that we do not find evidence for logarithmic growth of $q$, we impose $q_1 = 0$ and find $q_0 = 1.02 \pm 0.04$, which is interestingly consistent with the scale invariant spectrum $q_0 = 1$, suggested in~\cite{Harari:1987ht}, to within $\sim$5\%.  An additional argument in favor of $q_0 = 1$ is that the string loops appear logarithmically distributed in size, as shown in Fig.~\ref{fig: xi sub H} and as expected for a network of intersecting strings (see Methods).   

One difference between~\cite{Gorghetto:2020qws} and this work that may contribute to the difference in $q$ is that Ref.~\cite{Gorghetto:2020qws}
used $x_{\rm UV} = 4$; in Supp. Fig.~\ref{fig:UV_Fit_Variations} we show that using $x_{\rm UV} = 4$ in our fits also leads to positive $q_1$ at non-trivial significance (see Supp. Tab.~\ref{tab:uv_variations}); however, as illustrated in Supp. Fig.~\ref{fig:Large_UV_Cutoff} at large $\log m_r / H$ and $x_{\rm UV} = 4$ the fits become visibly poor at large $k/H$ because the spectrum $F(k/H)$ begins to drop rapidly for $k \sim m_r$.  The fact that~\cite{Gorghetto:2020qws} is only resolving the string cores by around one grid site at large $\log m_r / H$ may also play a role.  We test the importance of the string-core resolution by performing an alternate simulation where we do not add extra refinement levels after $\log m_r / H \approx 5.3$, such that $\Gamma$ is resolved by one grid site at $\log m_r / H \approx 8.1$ (see Supp. Fig.~\ref{fig:resolution}).  As illustrated in Supp. Fig.~\ref{fig:LoRes_Comparison}, in this case the spectrum becomes distinctly biased towards larger $q$ at larger $\log$, where the string-core resolution is low.   

Our result that $q_1$ is consistent with zero is robust to changes to $x_{\rm UV}$ (Supp. Fig.~\ref{fig:UV_Fit_Variations} and Tab.~\ref{tab:uv_variations}), for $32 \geq x_{\rm UV} \gtrsim 8$, to $x_{\rm IR}$ (Supp. Fig.~\ref{fig:IR_Fit_Variations} and Tab.~\ref{tab:ir_variations}), for the range $30 \leq x_{\rm IR} \leq 100$ that we consider, to the $\Delta \log$ size used in computing $F$ (Supp. Fig.~\ref{fig:Log_Fit_Variations} and Tab.~\ref{tab:deltaLog_variations}), for $0.125 \leq \Delta \log \leq \log 2$, and to the method used for regulating the string cores when computing $F$ (Supp. Fig.~\ref{fig:Masking_Fit_Variations} and Tab.~\ref{tab:masking_variations}).

\section*{Dark matter density}

The axion EOM during the QCD epoch generically violates number density conservation.  In particular, the non-linear axion potential is a function of $\cos(a/f_a)$, which implies that non-linear terms in the EOM are important if $|a / f_a| \gtrsim \pi$.  Given the instantaneous spectrum $F(k/H)$ we may compute the average field value squared at a given time $t$ by \mbox{$\langle (a / f_a)^2 \rangle \approx 4 \pi \int^t (dt'/t)  \xi(t') \langle (H'/k')^2 \rangle \log m_r / H'$}, with $\langle (H'/k')^2 \rangle$ being the expected value of $H/k$ at time $t'$ computed from the distribution $F(k/H)$ (see Methods and note that this is accurate to leading order in $\log m_r / H$).  We expect $\langle (H  / k)^2 \rangle$ to be proportional to $H^2 / k_{\rm IR}^2$, with $k_{\rm IR} / H \propto \sqrt{\xi}$ being the effective IR cut-off for $F(k/H)$ that arises from the typical separation of strings $\sim$$k_{\rm IR}^{-1}$; note that this implies that as $\xi(t)$ grows with time, the effective IR cut-off moves towards the UV like $\sqrt{\xi}$ because the strings become more closely packed together.  Let us define a dimensionless coefficient $\beta$ by the relation \mbox{$\langle (H / k)^2 \rangle^{-1} = \beta \, \xi$}; a fit of this functional form to the spectral data leads to $\beta = 840 \pm 70$ for $q = 1.06$ (see Supp. Fig.~\ref{fig:H_over_k2}).  Note that smaller values of $q$ lead to larger values of $\beta$ and that $q = 1.06$  is the maximum value of $q$ allowed at 1$\sigma$ from our analysis.
In terms of this coefficient $\langle (a / f_a)^2 \rangle \approx (4 \pi / \beta) \log m_r / H \lesssim 1.1$ (for $\log m_r / H \lesssim 70$), which implies that non-linear number changing processes are at most marginally relevant.  (Non-linear corrections to the linearized force are at most $\sim$15\%.)  This justifies our use of number density conservation below in estimating the DM abundance. 

\begin{figure}[htb]
\centering{
\includegraphics[width =.49 \textwidth]{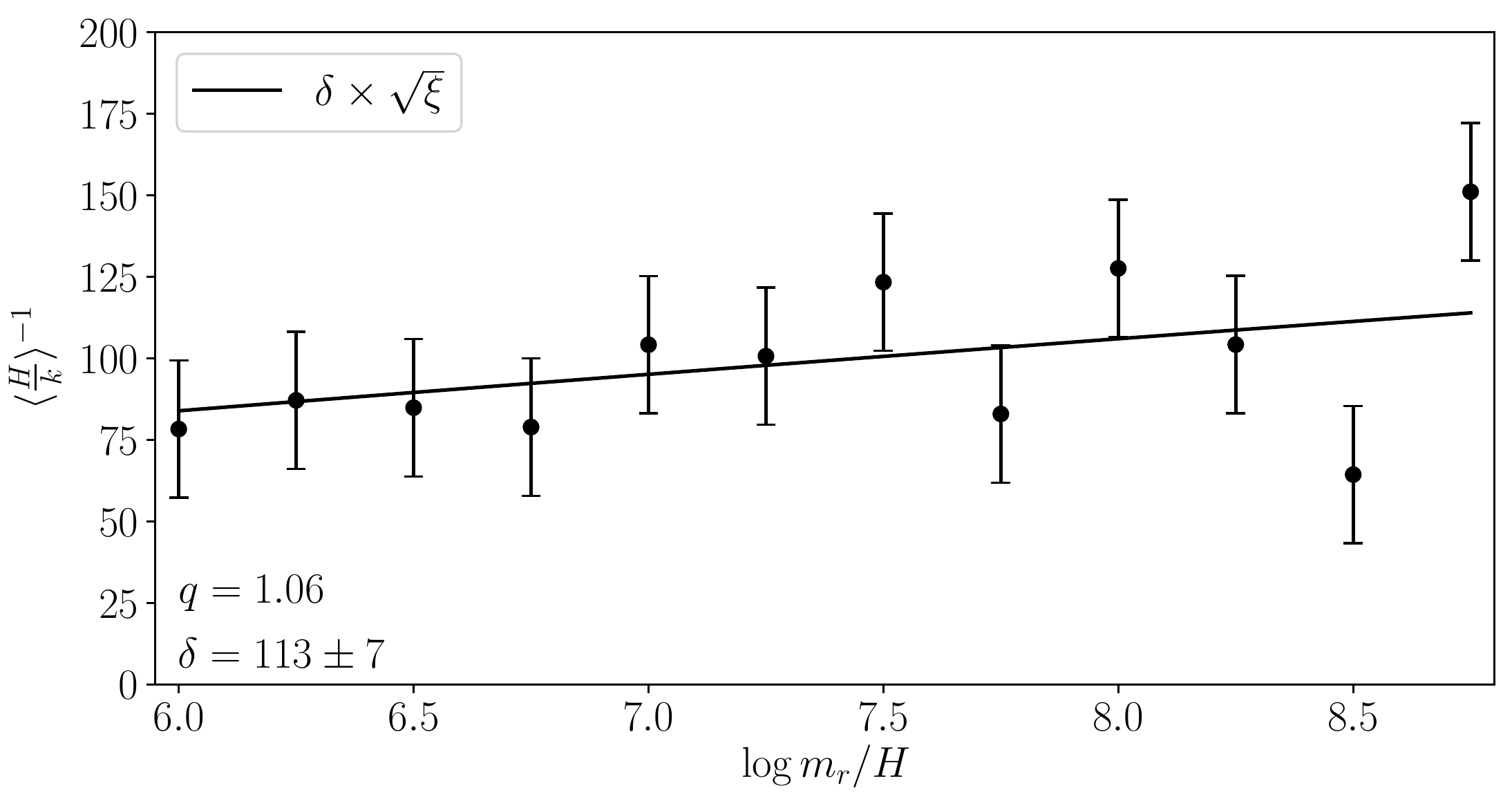}}
\caption{The inverse expectation value $\langle H / k \rangle^{-1}$ is computed using the instantaneous axion spectrum $F(k/H)$ by numerically integrating the spectrum to $k/H = x_{\rm max} = 50$ and then analytically integrating the power law distribution $F(x) \propto x^{-q}$ from $x_{\rm max}$ to the UV cut-off at $k/H \sim e^{\log_*}$ for $\log_* \approx 65$.  For $q > 1$ the expectation value does not strongly depend on the UV cut off but is instead a function of the effective IR cut-off, which is set by $\xi$ such that $\langle H/K \rangle^{-1} = \delta \sqrt{\xi}$ for some parameter $\delta$, which we determine by fitting this model to the numerical data as illustrated here.  Smaller values of $\delta$ correspond to larger axion number densities and thus large axion DM densities.  Here, we illustrate the result for the maximum allowed $q$ of $1.06$, which leads to the smallest $\delta$ consistent with our simulation results.  }
\label{fig:delta}
\end{figure}

To compute the axion number density we need to compute the expectation value $\langle H / k \rangle$ over the distribution $F(k/H)$.  Following the justification in the previous paragraph we may parameterize this expectation value in terms of the IR cut-off and thus $\xi$, $\langle H / k \rangle^{-1} = \delta \sqrt{\xi}$, for a dimensionless parameter $\delta$.  In Fig.~\ref{fig:delta} we illustrate the $\langle H / k \rangle^{-1}$ data, assuming $q = 1.06$, as a function of $\log m_r / H$ along with the best fit model, which leads to $\delta = 113 \pm 7$; note that smaller values of $q$ lead to larger values of $\delta$.  To compute $\langle H / k \rangle^{-1}$ (and also $\langle (H / k)^2 \rangle^{-1}$) we numerically integrate the spectrum up to $k / H = x_{\rm max}$, with $x_{\rm max} = 50$, and then analytically integrate the power-law functional form $F(k/H) \propto 1/k^q$ from $x_{\rm max}$ to $k/H \sim e^{\log_*}$, with $\log_* \sim 60 - 70$. 
The axion number density at the epoch of the QCD phase transition is then, to leading order in $\log_*$, $n_a^{\rm string} \approx (8 \pi f_a^2 H / \delta) \sqrt{\xi_*} \log_*$.  

If the spectrum is exactly scale invariant at large $k$, such that $q = 1$, then $\delta \propto \log(m_r / H)$.  Defining $\delta = \delta_1 \log(m_r / H)$ in this case we compute $\delta_1 = 6.2 \pm 0.4$.  The axion number density from strings is then $n_a^{\rm string} \approx (8 \pi f_a^2 H / \delta_1) \sqrt{\xi_*}$.  At 1$\sigma$ we find that $q$ could be as low as $q \approx 0.98$.  For $q < 1$ the quantity $\delta$ increases for increasing UV cut-offs like $(m_r / H)^{1-q}$; in particular, for $q = 0.98$ and $\log m_r / H = 70$ we calculate $\delta = 820 \pm 50$.  Thus, accounting for the uncertainty on $q$ from our simulations we find that $\delta$ is in the range $\delta \in (106,870)$.

Let us more precisely define the time $t_*$ as the time when the axion field becomes dynamical, which is when $3 H(t_*) = m_a(t_*)$, for a time-dependent mass $m_a(t)$ that is increasing rapidly during the QCD phase transition~\cite{diCortona:2015ldu}. The axion string network is observed to collapse around $t_*$ (see, {\it e.g.},~\cite{Buschmann:2019icd}), meaning that at times $t \gtrsim t_*$ axion number density is conserved.  Assuming axion number density conservation allows us to relate the present-day DM abundance to the expression for $n_a^{\rm string}$ at $t_*$ (see Methods):
\es{eq:omega_a_str}{
\Omega_a^{\rm str} \approx 0.12\, h^{-2} \left( {f_a \over 4.3 \cdot 10^{10}  {\rm GeV}} \right)^{1.17} {107 \over \delta} \sqrt{ {\xi_* \over 17}} {\log_* \over 70} . 
}
Axions produced from domain wall and misalignment dynamics during the QCD phase transition provide a sub-dominant contribution to the DM density~\cite{Buschmann:2019icd}: $\Omega_a^{\rm QCD} \approx 0.017\, h^{-2} (f_a / 4.3 \cdot 10^{10} \, {\rm GeV})^{1.17}$. (Note that we assume a domain wall number of unity so that domain walls are unstable, but see {\it e.g.}~\cite{Hiramatsu:2012sc}.)  The DM abundance as measured by the Planck Observatory using the cosmic microwave background is $\Omega_{\rm DM} = (0.12 \pm 0.0012) h^{-2}$, with $h$ the Hubble rate scaling factor~\cite{Aghanim:2018eyx}.  Adding in the contribution from the QCD phase transition $\Omega_a^{\rm QCD}$, and assuming $q \in (0.98,1.06)$, we find that the $f_a$ that gives rise to the observed DM abundance should be in the range $f_a \in (3.1 \times 10^{10}, 1.4 \times 10^{11})$ GeV ($m_a \in (40, 180)$ $\mu$eV), where for the lower $f_a$ bound we have conservatively allowed for the possibility that at $t_*$ the remaining energy density in strings is instantaneously deposited into axions with spectrum $F$, raising the string-induced DM density by a factor of $3/2$, though in actuality this contribution is likely smaller since the spectrum shifts towards the UV as $m_a(t)$ increases. If the index is scale invariant ($q = 1$), then we predict $m_a = 65 \pm 6$ $\mu$eV.

\section*{Discussion}

In this work we provide the largest and highest-resolution simulation of the axion string network to-date by making use of a novel AMR framework that allows us to resolve the axion string cores while maintaining lower resolution over the majority of the simulation volume.
Our AMR approach may be used in the future to simulate the axion dynamics at the QCD epoch where domain walls form and the string network collapses \cite{Buschmann:2019icd} and to study axion-like particle string networks that produce  gravitational wave radiation~\cite{Chang:2021afa, Figueroa:2020lvo, Gorghetto:2021fsn}.

Our results have important implications for axion direct detection experiments, as our preferred mass range of $(40,180)$ $\mu$eV is higher than that which may be probed by two of the main dedicated experiments that are aiming to test this cosmological scenario, {\it ADMX}~\cite{Braine:2019fqb} and {\it HAYSTAC}~\cite{HAYSTAC:2020kwv}. On the other hand, this mass range may be probed by {\it ADMX} with future searches \cite{Woollett:2018htk}, by the {\it MADMAX} experiment~\cite{MADMAX:2019pub, Beurthey:2020yuq}, and by the proposed plasma haloscope~\cite{Lawson:2019brd}. Our work motivates focusing experimental efforts on this mass range. 
The dominant source of uncertainty on $m_a$ in our estimates arises from the index $q$, which we find does not evolve with $\log m_r / H$ and is in the range $(0.98,1.06)$; this range is statistics-limited and will shrink with future simulation efforts using AMR, leading to more precise predictions that can in turn better inform experimental efforts.

\bibliography{Bibliography}

\begin{thebibliography}{53}%
\makeatletter
\providecommand \@ifxundefined [1]{%
 \@ifx{#1\undefined}
}%
\providecommand \@ifnum [1]{%
 \ifnum #1\expandafter \@firstoftwo
 \else \expandafter \@secondoftwo
 \fi
}%
\providecommand \@ifx [1]{%
 \ifx #1\expandafter \@firstoftwo
 \else \expandafter \@secondoftwo
 \fi
}%
\providecommand \natexlab [1]{#1}%
\providecommand \enquote  [1]{``#1''}%
\providecommand \bibnamefont  [1]{#1}%
\providecommand \bibfnamefont [1]{#1}%
\providecommand \citenamefont [1]{#1}%
\providecommand \href@noop [0]{\@secondoftwo}%
\providecommand \href [0]{\begingroup \@sanitize@url \@href}%
\providecommand \@href[1]{\@@startlink{#1}\@@href}%
\providecommand \@@href[1]{\endgroup#1\@@endlink}%
\providecommand \@sanitize@url [0]{\catcode `\\12\catcode `\$12\catcode
  `\&12\catcode `\#12\catcode `\^12\catcode `\_12\catcode `\%12\relax}%
\providecommand \@@startlink[1]{}%
\providecommand \@@endlink[0]{}%
\providecommand \url  [0]{\begingroup\@sanitize@url \@url }%
\providecommand \@url [1]{\endgroup\@href {#1}{\urlprefix }}%
\providecommand \urlprefix  [0]{URL }%
\providecommand \Eprint [0]{\href }%
\providecommand \doibase [0]{http://dx.doi.org/}%
\providecommand \selectlanguage [0]{\@gobble}%
\providecommand \bibinfo  [0]{\@secondoftwo}%
\providecommand \bibfield  [0]{\@secondoftwo}%
\providecommand \translation [1]{[#1]}%
\providecommand \BibitemOpen [0]{}%
\providecommand \bibitemStop [0]{}%
\providecommand \bibitemNoStop [0]{.\EOS\space}%
\providecommand \EOS [0]{\spacefactor3000\relax}%
\providecommand \BibitemShut  [1]{\csname bibitem#1\endcsname}%
\let\auto@bib@innerbib\@empty
\bibitem [{\citenamefont {Abel}\ \emph {et~al.}(2020)\citenamefont {Abel} \emph
  {et~al.}}]{nEDM:2020crw}%
  \BibitemOpen
  \bibfield  {author} {\bibinfo {author} {\bibfnamefont {C.}~\bibnamefont
  {Abel}} \emph {et~al.} (\bibinfo {collaboration} {nEDM}),\ }\bibfield
  {title} {\enquote {\bibinfo {title} {{Measurement of the permanent electric
  dipole moment of the neutron}},}\ }\href {\doibase
  10.1103/PhysRevLett.124.081803} {\bibfield  {journal} {\bibinfo  {journal}
  {Phys. Rev. Lett.}\ }\textbf {\bibinfo {volume} {124}},\ \bibinfo {pages}
  {081803} (\bibinfo {year} {2020})},\ \Eprint
  {http://arxiv.org/abs/2001.11966} {arXiv:2001.11966 [hep-ex]} \BibitemShut
  {NoStop}%
\bibitem [{\citenamefont {Peccei}\ and\ \citenamefont
  {Quinn}(1977{\natexlab{a}})}]{Peccei:1977hh}%
  \BibitemOpen
  \bibfield  {author} {\bibinfo {author} {\bibfnamefont {R.~D.}\ \bibnamefont
  {Peccei}}\ and\ \bibinfo {author} {\bibfnamefont {Helen~R.}\ \bibnamefont
  {Quinn}},\ }\bibfield  {title} {\enquote {\bibinfo {title} {{CP Conservation
  in the Presence of Instantons}},}\ }\href {\doibase
  10.1103/PhysRevLett.38.1440} {\bibfield  {journal} {\bibinfo  {journal}
  {Phys. Rev. Lett.}\ }\textbf {\bibinfo {volume} {38}},\ \bibinfo {pages}
  {1440--1443} (\bibinfo {year} {1977}{\natexlab{a}})}\BibitemShut {NoStop}%
\bibitem [{\citenamefont {Peccei}\ and\ \citenamefont
  {Quinn}(1977{\natexlab{b}})}]{Peccei:1977ur}%
  \BibitemOpen
  \bibfield  {author} {\bibinfo {author} {\bibfnamefont {R.~D.}\ \bibnamefont
  {Peccei}}\ and\ \bibinfo {author} {\bibfnamefont {Helen~R.}\ \bibnamefont
  {Quinn}},\ }\bibfield  {title} {\enquote {\bibinfo {title} {{Constraints
  Imposed by CP Conservation in the Presence of Instantons}},}\ }\href
  {\doibase 10.1103/PhysRevD.16.1791} {\bibfield  {journal} {\bibinfo
  {journal} {Phys. Rev.}\ }\textbf {\bibinfo {volume} {D16}},\ \bibinfo {pages}
  {1791--1797} (\bibinfo {year} {1977}{\natexlab{b}})}\BibitemShut {NoStop}%
\bibitem [{\citenamefont {Weinberg}(1978)}]{Weinberg:1977ma}%
  \BibitemOpen
  \bibfield  {author} {\bibinfo {author} {\bibfnamefont {Steven}\ \bibnamefont
  {Weinberg}},\ }\bibfield  {title} {\enquote {\bibinfo {title} {{A New Light
  Boson?}}}\ }\href {\doibase 10.1103/PhysRevLett.40.223} {\bibfield  {journal}
  {\bibinfo  {journal} {Phys. Rev. Lett.}\ }\textbf {\bibinfo {volume} {40}},\
  \bibinfo {pages} {223--226} (\bibinfo {year} {1978})}\BibitemShut {NoStop}%
\bibitem [{\citenamefont {Wilczek}(1978)}]{Wilczek:1977pj}%
  \BibitemOpen
  \bibfield  {author} {\bibinfo {author} {\bibfnamefont {Frank}\ \bibnamefont
  {Wilczek}},\ }\bibfield  {title} {\enquote {\bibinfo {title} {{Problem of
  Strong p and t Invariance in the Presence of Instantons}},}\ }\href {\doibase
  10.1103/PhysRevLett.40.279} {\bibfield  {journal} {\bibinfo  {journal} {Phys.
  Rev. Lett.}\ }\textbf {\bibinfo {volume} {40}},\ \bibinfo {pages} {279--282}
  (\bibinfo {year} {1978})}\BibitemShut {NoStop}%
\bibitem [{\citenamefont {Preskill}\ \emph {et~al.}(1983)\citenamefont
  {Preskill}, \citenamefont {Wise},\ and\ \citenamefont
  {Wilczek}}]{Preskill:1982cy}%
  \BibitemOpen
  \bibfield  {author} {\bibinfo {author} {\bibfnamefont {John}\ \bibnamefont
  {Preskill}}, \bibinfo {author} {\bibfnamefont {Mark~B.}\ \bibnamefont
  {Wise}}, \ and\ \bibinfo {author} {\bibfnamefont {Frank}\ \bibnamefont
  {Wilczek}},\ }\bibfield  {title} {\enquote {\bibinfo {title} {{Cosmology of
  the Invisible Axion}},}\ }\href {\doibase 10.1016/0370-2693(83)90637-8}
  {\bibfield  {journal} {\bibinfo  {journal} {Phys. Lett.}\ }\textbf {\bibinfo
  {volume} {120B}},\ \bibinfo {pages} {127--132} (\bibinfo {year}
  {1983})}\BibitemShut {NoStop}%
\bibitem [{\citenamefont {Abbott}\ and\ \citenamefont
  {Sikivie}(1983)}]{Abbott:1982af}%
  \BibitemOpen
  \bibfield  {author} {\bibinfo {author} {\bibfnamefont {L.~F.}\ \bibnamefont
  {Abbott}}\ and\ \bibinfo {author} {\bibfnamefont {P.}~\bibnamefont
  {Sikivie}},\ }\bibfield  {title} {\enquote {\bibinfo {title} {{A Cosmological
  Bound on the Invisible Axion}},}\ }\href {\doibase
  10.1016/0370-2693(83)90638-X} {\bibfield  {journal} {\bibinfo  {journal}
  {Phys. Lett.}\ }\textbf {\bibinfo {volume} {120B}},\ \bibinfo {pages}
  {133--136} (\bibinfo {year} {1983})}\BibitemShut {NoStop}%
\bibitem [{\citenamefont {Dine}\ and\ \citenamefont
  {Fischler}(1983)}]{Dine:1982ah}%
  \BibitemOpen
  \bibfield  {author} {\bibinfo {author} {\bibfnamefont {Michael}\ \bibnamefont
  {Dine}}\ and\ \bibinfo {author} {\bibfnamefont {Willy}\ \bibnamefont
  {Fischler}},\ }\bibfield  {title} {\enquote {\bibinfo {title} {{The Not So
  Harmless Axion}},}\ }\href {\doibase 10.1016/0370-2693(83)90639-1} {\bibfield
   {journal} {\bibinfo  {journal} {Phys. Lett.}\ }\textbf {\bibinfo {volume}
  {120B}},\ \bibinfo {pages} {137--141} (\bibinfo {year} {1983})}\BibitemShut
  {NoStop}%
\bibitem [{\citenamefont {Graham}\ \emph {et~al.}(2015)\citenamefont {Graham},
  \citenamefont {Irastorza}, \citenamefont {Lamoreaux}, \citenamefont
  {Lindner},\ and\ \citenamefont {van Bibber}}]{Graham:2015ouw}%
  \BibitemOpen
  \bibfield  {author} {\bibinfo {author} {\bibfnamefont {Peter~W.}\
  \bibnamefont {Graham}}, \bibinfo {author} {\bibfnamefont {Igor~G.}\
  \bibnamefont {Irastorza}}, \bibinfo {author} {\bibfnamefont {Steven~K.}\
  \bibnamefont {Lamoreaux}}, \bibinfo {author} {\bibfnamefont {Axel}\
  \bibnamefont {Lindner}}, \ and\ \bibinfo {author} {\bibfnamefont {Karl~A.}\
  \bibnamefont {van Bibber}},\ }\bibfield  {title} {\enquote {\bibinfo {title}
  {{Experimental Searches for the Axion and Axion-Like Particles}},}\ }\href
  {\doibase 10.1146/annurev-nucl-102014-022120} {\bibfield  {journal} {\bibinfo
   {journal} {Ann. Rev. Nucl. Part. Sci.}\ }\textbf {\bibinfo {volume} {65}},\
  \bibinfo {pages} {485--514} (\bibinfo {year} {2015})},\ \Eprint
  {http://arxiv.org/abs/1602.00039} {arXiv:1602.00039 [hep-ex]} \BibitemShut
  {NoStop}%
\bibitem [{\citenamefont {Sikivie}(2021)}]{Sikivie:2020zpn}%
  \BibitemOpen
  \bibfield  {author} {\bibinfo {author} {\bibfnamefont {Pierre}\ \bibnamefont
  {Sikivie}},\ }\bibfield  {title} {\enquote {\bibinfo {title} {{Invisible
  Axion Search Methods}},}\ }\href {\doibase 10.1103/RevModPhys.93.015004}
  {\bibfield  {journal} {\bibinfo  {journal} {Rev. Mod. Phys.}\ }\textbf
  {\bibinfo {volume} {93}},\ \bibinfo {pages} {015004} (\bibinfo {year}
  {2021})},\ \Eprint {http://arxiv.org/abs/2003.02206} {arXiv:2003.02206
  [hep-ph]} \BibitemShut {NoStop}%
\bibitem [{\citenamefont {Di~Luzio}\ \emph {et~al.}(2020)\citenamefont
  {Di~Luzio}, \citenamefont {Giannotti}, \citenamefont {Nardi},\ and\
  \citenamefont {Visinelli}}]{DiLuzio:2020wdo}%
  \BibitemOpen
  \bibfield  {author} {\bibinfo {author} {\bibfnamefont {Luca}\ \bibnamefont
  {Di~Luzio}}, \bibinfo {author} {\bibfnamefont {Maurizio}\ \bibnamefont
  {Giannotti}}, \bibinfo {author} {\bibfnamefont {Enrico}\ \bibnamefont
  {Nardi}}, \ and\ \bibinfo {author} {\bibfnamefont {Luca}\ \bibnamefont
  {Visinelli}},\ }\bibfield  {title} {\enquote {\bibinfo {title} {{The
  landscape of QCD axion models}},}\ }\href {\doibase
  10.1016/j.physrep.2020.06.002} {\bibfield  {journal} {\bibinfo  {journal}
  {Phys. Rept.}\ }\textbf {\bibinfo {volume} {870}},\ \bibinfo {pages} {1--117}
  (\bibinfo {year} {2020})},\ \Eprint {http://arxiv.org/abs/2003.01100}
  {arXiv:2003.01100 [hep-ph]} \BibitemShut {NoStop}%
\bibitem [{\citenamefont {Marsh}(2016)}]{Marsh:2015xka}%
  \BibitemOpen
  \bibfield  {author} {\bibinfo {author} {\bibfnamefont {David J.~E.}\
  \bibnamefont {Marsh}},\ }\bibfield  {title} {\enquote {\bibinfo {title}
  {{Axion Cosmology}},}\ }\href {\doibase 10.1016/j.physrep.2016.06.005}
  {\bibfield  {journal} {\bibinfo  {journal} {Phys. Rept.}\ }\textbf {\bibinfo
  {volume} {643}},\ \bibinfo {pages} {1--79} (\bibinfo {year} {2016})},\
  \Eprint {http://arxiv.org/abs/1510.07633} {arXiv:1510.07633 [astro-ph.CO]}
  \BibitemShut {NoStop}%
\bibitem [{\citenamefont {Harari}\ and\ \citenamefont
  {Sikivie}(1987)}]{Harari:1987ht}%
  \BibitemOpen
  \bibfield  {author} {\bibinfo {author} {\bibfnamefont {Diego}\ \bibnamefont
  {Harari}}\ and\ \bibinfo {author} {\bibfnamefont {P.}~\bibnamefont
  {Sikivie}},\ }\bibfield  {title} {\enquote {\bibinfo {title} {{On the
  Evolution of Global Strings in the Early Universe}},}\ }\href {\doibase
  10.1016/0370-2693(87)90032-3} {\bibfield  {journal} {\bibinfo  {journal}
  {Phys. Lett. B}\ }\textbf {\bibinfo {volume} {195}},\ \bibinfo {pages}
  {361--365} (\bibinfo {year} {1987})}\BibitemShut {NoStop}%
\bibitem [{\citenamefont {Hagmann}\ and\ \citenamefont
  {Sikivie}(1991)}]{Hagmann:1990mj}%
  \BibitemOpen
  \bibfield  {author} {\bibinfo {author} {\bibfnamefont {C.}~\bibnamefont
  {Hagmann}}\ and\ \bibinfo {author} {\bibfnamefont {P.}~\bibnamefont
  {Sikivie}},\ }\bibfield  {title} {\enquote {\bibinfo {title} {{Computer
  simulations of the motion and decay of global strings}},}\ }\href {\doibase
  10.1016/0550-3213(91)90243-Q} {\bibfield  {journal} {\bibinfo  {journal}
  {Nucl. Phys. B}\ }\textbf {\bibinfo {volume} {363}},\ \bibinfo {pages}
  {247--280} (\bibinfo {year} {1991})}\BibitemShut {NoStop}%
\bibitem [{\citenamefont {Davis}\ and\ \citenamefont
  {Shellard}(1989)}]{Davis:1989nj}%
  \BibitemOpen
  \bibfield  {author} {\bibinfo {author} {\bibfnamefont {R.~L.}\ \bibnamefont
  {Davis}}\ and\ \bibinfo {author} {\bibfnamefont {E.~P.~S.}\ \bibnamefont
  {Shellard}},\ }\bibfield  {title} {\enquote {\bibinfo {title} {{Do Axions
  Need Inflation?}}}\ }\href {\doibase 10.1016/0550-3213(89)90187-9} {\bibfield
   {journal} {\bibinfo  {journal} {Nucl. Phys. B}\ }\textbf {\bibinfo {volume}
  {324}},\ \bibinfo {pages} {167--186} (\bibinfo {year} {1989})}\BibitemShut
  {NoStop}%
\bibitem [{\citenamefont {Battye}\ and\ \citenamefont
  {Shellard}(1994{\natexlab{a}})}]{Battye:1993jv}%
  \BibitemOpen
  \bibfield  {author} {\bibinfo {author} {\bibfnamefont {R.~A.}\ \bibnamefont
  {Battye}}\ and\ \bibinfo {author} {\bibfnamefont {E.~P.~S.}\ \bibnamefont
  {Shellard}},\ }\bibfield  {title} {\enquote {\bibinfo {title} {{Global string
  radiation}},}\ }\href {\doibase 10.1016/0550-3213(94)90573-8} {\bibfield
  {journal} {\bibinfo  {journal} {Nucl. Phys. B}\ }\textbf {\bibinfo {volume}
  {423}},\ \bibinfo {pages} {260--304} (\bibinfo {year}
  {1994}{\natexlab{a}})},\ \Eprint {http://arxiv.org/abs/astro-ph/9311017}
  {arXiv:astro-ph/9311017} \BibitemShut {NoStop}%
\bibitem [{\citenamefont {Battye}\ and\ \citenamefont
  {Shellard}(1994{\natexlab{b}})}]{Battye:1994au}%
  \BibitemOpen
  \bibfield  {author} {\bibinfo {author} {\bibfnamefont {R.~A.}\ \bibnamefont
  {Battye}}\ and\ \bibinfo {author} {\bibfnamefont {E.~P.~S.}\ \bibnamefont
  {Shellard}},\ }\bibfield  {title} {\enquote {\bibinfo {title} {{Axion string
  constraints}},}\ }\href {\doibase 10.1103/PhysRevLett.73.2954} {\bibfield
  {journal} {\bibinfo  {journal} {Phys. Rev. Lett.}\ }\textbf {\bibinfo
  {volume} {73}},\ \bibinfo {pages} {2954--2957} (\bibinfo {year}
  {1994}{\natexlab{b}})},\ \bibinfo {note} {[Erratum: Phys.Rev.Lett. 76,
  2203--2204 (1996)]},\ \Eprint {http://arxiv.org/abs/astro-ph/9403018}
  {arXiv:astro-ph/9403018} \BibitemShut {NoStop}%
\bibitem [{\citenamefont {Vilenkin}\ and\ \citenamefont
  {Everett}(1982)}]{Vilenkin:1982ks}%
  \BibitemOpen
  \bibfield  {author} {\bibinfo {author} {\bibfnamefont {A.}~\bibnamefont
  {Vilenkin}}\ and\ \bibinfo {author} {\bibfnamefont {A.~E.}\ \bibnamefont
  {Everett}},\ }\bibfield  {title} {\enquote {\bibinfo {title} {{Cosmic Strings
  and Domain Walls in Models with Goldstone and PseudoGoldstone Bosons}},}\
  }\href {\doibase 10.1103/PhysRevLett.48.1867} {\bibfield  {journal} {\bibinfo
   {journal} {Phys. Rev. Lett.}\ }\textbf {\bibinfo {volume} {48}},\ \bibinfo
  {pages} {1867--1870} (\bibinfo {year} {1982})}\BibitemShut {NoStop}%
\bibitem [{\citenamefont {Sikivie}(1982)}]{Sikivie:1982qv}%
  \BibitemOpen
  \bibfield  {author} {\bibinfo {author} {\bibfnamefont {P.}~\bibnamefont
  {Sikivie}},\ }\bibfield  {title} {\enquote {\bibinfo {title} {{Of Axions,
  Domain Walls and the Early Universe}},}\ }\href {\doibase
  10.1103/PhysRevLett.48.1156} {\bibfield  {journal} {\bibinfo  {journal}
  {Phys. Rev. Lett.}\ }\textbf {\bibinfo {volume} {48}},\ \bibinfo {pages}
  {1156--1159} (\bibinfo {year} {1982})}\BibitemShut {NoStop}%
\bibitem [{\citenamefont {Davis}(1986)}]{Davis:1986xc}%
  \BibitemOpen
  \bibfield  {author} {\bibinfo {author} {\bibfnamefont {Richard~Lynn}\
  \bibnamefont {Davis}},\ }\bibfield  {title} {\enquote {\bibinfo {title}
  {{Cosmic Axions from Cosmic Strings}},}\ }\href {\doibase
  10.1016/0370-2693(86)90300-X} {\bibfield  {journal} {\bibinfo  {journal}
  {Phys. Lett. B}\ }\textbf {\bibinfo {volume} {180}},\ \bibinfo {pages}
  {225--230} (\bibinfo {year} {1986})}\BibitemShut {NoStop}%
\bibitem [{\citenamefont {Shellard}(1987)}]{Shellard:1987bv}%
  \BibitemOpen
  \bibfield  {author} {\bibinfo {author} {\bibfnamefont {E.~P.~S.}\
  \bibnamefont {Shellard}},\ }\bibfield  {title} {\enquote {\bibinfo {title}
  {{Cosmic String Interactions}},}\ }\href {\doibase
  10.1016/0550-3213(87)90290-2} {\bibfield  {journal} {\bibinfo  {journal}
  {Nucl. Phys. B}\ }\textbf {\bibinfo {volume} {283}},\ \bibinfo {pages}
  {624--656} (\bibinfo {year} {1987})}\BibitemShut {NoStop}%
\bibitem [{\citenamefont {Yamaguchi}\ \emph {et~al.}(1999)\citenamefont
  {Yamaguchi}, \citenamefont {Kawasaki},\ and\ \citenamefont
  {Yokoyama}}]{Yamaguchi:1998gx}%
  \BibitemOpen
  \bibfield  {author} {\bibinfo {author} {\bibfnamefont {Masahide}\
  \bibnamefont {Yamaguchi}}, \bibinfo {author} {\bibfnamefont {M.}~\bibnamefont
  {Kawasaki}}, \ and\ \bibinfo {author} {\bibfnamefont {Jun'ichi}\ \bibnamefont
  {Yokoyama}},\ }\bibfield  {title} {\enquote {\bibinfo {title} {{Evolution of
  axionic strings and spectrum of axions radiated from them}},}\ }\href
  {\doibase 10.1103/PhysRevLett.82.4578} {\bibfield  {journal} {\bibinfo
  {journal} {Phys. Rev. Lett.}\ }\textbf {\bibinfo {volume} {82}},\ \bibinfo
  {pages} {4578--4581} (\bibinfo {year} {1999})},\ \Eprint
  {http://arxiv.org/abs/hep-ph/9811311} {arXiv:hep-ph/9811311} \BibitemShut
  {NoStop}%
\bibitem [{\citenamefont {Klaer}\ and\ \citenamefont
  {Moore}(2017)}]{Klaer:2017ond}%
  \BibitemOpen
  \bibfield  {author} {\bibinfo {author} {\bibfnamefont {Vincent B.~.}\
  \bibnamefont {Klaer}}\ and\ \bibinfo {author} {\bibfnamefont {Guy~D.}\
  \bibnamefont {Moore}},\ }\bibfield  {title} {\enquote {\bibinfo {title} {{The
  dark-matter axion mass}},}\ }\href {\doibase 10.1088/1475-7516/2017/11/049}
  {\bibfield  {journal} {\bibinfo  {journal} {JCAP}\ }\textbf {\bibinfo
  {volume} {11}},\ \bibinfo {pages} {049} (\bibinfo {year} {2017})},\ \Eprint
  {http://arxiv.org/abs/1708.07521} {arXiv:1708.07521 [hep-ph]} \BibitemShut
  {NoStop}%
\bibitem [{\citenamefont {Gorghetto}\ \emph {et~al.}(2018)\citenamefont
  {Gorghetto}, \citenamefont {Hardy},\ and\ \citenamefont
  {Villadoro}}]{Gorghetto:2018myk}%
  \BibitemOpen
  \bibfield  {author} {\bibinfo {author} {\bibfnamefont {Marco}\ \bibnamefont
  {Gorghetto}}, \bibinfo {author} {\bibfnamefont {Edward}\ \bibnamefont
  {Hardy}}, \ and\ \bibinfo {author} {\bibfnamefont {Giovanni}\ \bibnamefont
  {Villadoro}},\ }\bibfield  {title} {\enquote {\bibinfo {title} {{Axions from
  Strings: the Attractive Solution}},}\ }\href {\doibase
  10.1007/JHEP07(2018)151} {\bibfield  {journal} {\bibinfo  {journal} {JHEP}\
  }\textbf {\bibinfo {volume} {07}},\ \bibinfo {pages} {151} (\bibinfo {year}
  {2018})},\ \Eprint {http://arxiv.org/abs/1806.04677} {arXiv:1806.04677
  [hep-ph]} \BibitemShut {NoStop}%
\bibitem [{\citenamefont {Vaquero}\ \emph {et~al.}(2019)\citenamefont
  {Vaquero}, \citenamefont {Redondo},\ and\ \citenamefont
  {Stadler}}]{Vaquero:2018tib}%
  \BibitemOpen
  \bibfield  {author} {\bibinfo {author} {\bibfnamefont {Alejandro}\
  \bibnamefont {Vaquero}}, \bibinfo {author} {\bibfnamefont {Javier}\
  \bibnamefont {Redondo}}, \ and\ \bibinfo {author} {\bibfnamefont {Julia}\
  \bibnamefont {Stadler}},\ }\bibfield  {title} {\enquote {\bibinfo {title}
  {{Early seeds of axion miniclusters}},}\ }\href {\doibase
  10.1088/1475-7516/2019/04/012} {\bibfield  {journal} {\bibinfo  {journal}
  {JCAP}\ }\textbf {\bibinfo {volume} {04}},\ \bibinfo {pages} {012} (\bibinfo
  {year} {2019})},\ \Eprint {http://arxiv.org/abs/1809.09241} {arXiv:1809.09241
  [astro-ph.CO]} \BibitemShut {NoStop}%
\bibitem [{\citenamefont {Buschmann}\ \emph {et~al.}(2020)\citenamefont
  {Buschmann}, \citenamefont {Foster},\ and\ \citenamefont
  {Safdi}}]{Buschmann:2019icd}%
  \BibitemOpen
  \bibfield  {author} {\bibinfo {author} {\bibfnamefont {Malte}\ \bibnamefont
  {Buschmann}}, \bibinfo {author} {\bibfnamefont {Joshua~W.}\ \bibnamefont
  {Foster}}, \ and\ \bibinfo {author} {\bibfnamefont {Benjamin~R.}\
  \bibnamefont {Safdi}},\ }\bibfield  {title} {\enquote {\bibinfo {title}
  {{Early-Universe Simulations of the Cosmological Axion}},}\ }\href {\doibase
  10.1103/PhysRevLett.124.161103} {\bibfield  {journal} {\bibinfo  {journal}
  {Phys. Rev. Lett.}\ }\textbf {\bibinfo {volume} {124}},\ \bibinfo {pages}
  {161103} (\bibinfo {year} {2020})},\ \Eprint
  {http://arxiv.org/abs/1906.00967} {arXiv:1906.00967 [astro-ph.CO]}
  \BibitemShut {NoStop}%
\bibitem [{\citenamefont {Gorghetto}\ \emph
  {et~al.}(2021{\natexlab{a}})\citenamefont {Gorghetto}, \citenamefont
  {Hardy},\ and\ \citenamefont {Villadoro}}]{Gorghetto:2020qws}%
  \BibitemOpen
  \bibfield  {author} {\bibinfo {author} {\bibfnamefont {Marco}\ \bibnamefont
  {Gorghetto}}, \bibinfo {author} {\bibfnamefont {Edward}\ \bibnamefont
  {Hardy}}, \ and\ \bibinfo {author} {\bibfnamefont {Giovanni}\ \bibnamefont
  {Villadoro}},\ }\bibfield  {title} {\enquote {\bibinfo {title} {{More Axions
  from Strings}},}\ }\href {\doibase 10.21468/SciPostPhys.10.2.050} {\bibfield
  {journal} {\bibinfo  {journal} {SciPost Phys.}\ }\textbf {\bibinfo {volume}
  {10}},\ \bibinfo {pages} {050} (\bibinfo {year} {2021}{\natexlab{a}})},\
  \Eprint {http://arxiv.org/abs/2007.04990} {arXiv:2007.04990 [hep-ph]}
  \BibitemShut {NoStop}%
\bibitem [{\citenamefont {Dine}\ \emph {et~al.}(2020)\citenamefont {Dine},
  \citenamefont {Fernandez}, \citenamefont {Ghalsasi},\ and\ \citenamefont
  {Patel}}]{Dine:2020pds}%
  \BibitemOpen
  \bibfield  {author} {\bibinfo {author} {\bibfnamefont {Michael}\ \bibnamefont
  {Dine}}, \bibinfo {author} {\bibfnamefont {Nicolas}\ \bibnamefont
  {Fernandez}}, \bibinfo {author} {\bibfnamefont {Akshay}\ \bibnamefont
  {Ghalsasi}}, \ and\ \bibinfo {author} {\bibfnamefont {Hiren~H.}\ \bibnamefont
  {Patel}},\ }\bibfield  {title} {\enquote {\bibinfo {title} {{Comments on
  Axions, Domain Walls, and Cosmic Strings}},}\ }\href@noop {} {\  (\bibinfo
  {year} {2020})},\ \Eprint {http://arxiv.org/abs/2012.13065} {arXiv:2012.13065
  [hep-ph]} \BibitemShut {NoStop}%
\bibitem [{\citenamefont {Drew}\ and\ \citenamefont
  {Shellard}(2019)}]{Drew:2019mzc}%
  \BibitemOpen
  \bibfield  {author} {\bibinfo {author} {\bibfnamefont {Amelia}\ \bibnamefont
  {Drew}}\ and\ \bibinfo {author} {\bibfnamefont {E.~P.~S.}\ \bibnamefont
  {Shellard}},\ }\bibfield  {title} {\enquote {\bibinfo {title} {{Radiation
  from Global Topological Strings using Adaptive Mesh Refinement: Methodology
  and Massless Modes}},}\ }\href@noop {} {\  (\bibinfo {year} {2019})},\
  \Eprint {http://arxiv.org/abs/1910.01718} {arXiv:1910.01718 [astro-ph.CO]}
  \BibitemShut {NoStop}%
\bibitem [{\citenamefont {Grilli~di Cortona}\ \emph {et~al.}(2016)\citenamefont
  {Grilli~di Cortona}, \citenamefont {Hardy}, \citenamefont {Pardo~Vega},\ and\
  \citenamefont {Villadoro}}]{diCortona:2015ldu}%
  \BibitemOpen
  \bibfield  {author} {\bibinfo {author} {\bibfnamefont {Giovanni}\
  \bibnamefont {Grilli~di Cortona}}, \bibinfo {author} {\bibfnamefont {Edward}\
  \bibnamefont {Hardy}}, \bibinfo {author} {\bibfnamefont {Javier}\
  \bibnamefont {Pardo~Vega}}, \ and\ \bibinfo {author} {\bibfnamefont
  {Giovanni}\ \bibnamefont {Villadoro}},\ }\bibfield  {title} {\enquote
  {\bibinfo {title} {{The QCD axion, precisely}},}\ }\href {\doibase
  10.1007/JHEP01(2016)034} {\bibfield  {journal} {\bibinfo  {journal} {JHEP}\
  }\textbf {\bibinfo {volume} {01}},\ \bibinfo {pages} {034} (\bibinfo {year}
  {2016})},\ \Eprint {http://arxiv.org/abs/1511.02867} {arXiv:1511.02867
  [hep-ph]} \BibitemShut {NoStop}%
\bibitem [{\citenamefont {Zhang}\ \emph {et~al.}(2020)\citenamefont {Zhang},
  \citenamefont {Myers}, \citenamefont {Gott}, \citenamefont {Almgren},\ and\
  \citenamefont {Bell}}]{zhang2020amrex}%
  \BibitemOpen
  \bibfield  {author} {\bibinfo {author} {\bibfnamefont {Weiqun}\ \bibnamefont
  {Zhang}}, \bibinfo {author} {\bibfnamefont {Andrew}\ \bibnamefont {Myers}},
  \bibinfo {author} {\bibfnamefont {Kevin}\ \bibnamefont {Gott}}, \bibinfo
  {author} {\bibfnamefont {Ann}\ \bibnamefont {Almgren}}, \ and\ \bibinfo
  {author} {\bibfnamefont {John}\ \bibnamefont {Bell}},\ }\href@noop {}
  {\enquote {\bibinfo {title} {Amrex: Block-structured adaptive mesh refinement
  for multiphysics applications},}\ } (\bibinfo {year} {2020}),\ \Eprint
  {http://arxiv.org/abs/2009.12009} {arXiv:2009.12009 [cs.MS]} \BibitemShut
  {NoStop}%
\bibitem [{\citenamefont {Hiramatsu}\ \emph {et~al.}(2012)\citenamefont
  {Hiramatsu}, \citenamefont {Kawasaki}, \citenamefont {Saikawa},\ and\
  \citenamefont {Sekiguchi}}]{Hiramatsu:2012gg}%
  \BibitemOpen
  \bibfield  {author} {\bibinfo {author} {\bibfnamefont {Takashi}\ \bibnamefont
  {Hiramatsu}}, \bibinfo {author} {\bibfnamefont {Masahiro}\ \bibnamefont
  {Kawasaki}}, \bibinfo {author} {\bibfnamefont {Ken'ichi}\ \bibnamefont
  {Saikawa}}, \ and\ \bibinfo {author} {\bibfnamefont {Toyokazu}\ \bibnamefont
  {Sekiguchi}},\ }\bibfield  {title} {\enquote {\bibinfo {title} {{Production
  of dark matter axions from collapse of string-wall systems}},}\ }\href
  {\doibase 10.1103/PhysRevD.85.105020} {\bibfield  {journal} {\bibinfo
  {journal} {Phys. Rev. D}\ }\textbf {\bibinfo {volume} {85}},\ \bibinfo
  {pages} {105020} (\bibinfo {year} {2012})},\ \bibinfo {note} {[Erratum:
  Phys.Rev.D 86, 089902 (2012)]},\ \Eprint {http://arxiv.org/abs/1202.5851}
  {arXiv:1202.5851 [hep-ph]} \BibitemShut {NoStop}%
\bibitem [{\citenamefont {Fleury}\ and\ \citenamefont
  {Moore}(2016)}]{Fleury:2015aca}%
  \BibitemOpen
  \bibfield  {author} {\bibinfo {author} {\bibfnamefont {Leesa}\ \bibnamefont
  {Fleury}}\ and\ \bibinfo {author} {\bibfnamefont {Guy~D.}\ \bibnamefont
  {Moore}},\ }\bibfield  {title} {\enquote {\bibinfo {title} {{Axion dark
  matter: strings and their cores}},}\ }\href {\doibase
  10.1088/1475-7516/2016/01/004} {\bibfield  {journal} {\bibinfo  {journal}
  {JCAP}\ }\textbf {\bibinfo {volume} {01}},\ \bibinfo {pages} {004} (\bibinfo
  {year} {2016})},\ \Eprint {http://arxiv.org/abs/1509.00026} {arXiv:1509.00026
  [hep-ph]} \BibitemShut {NoStop}%
\bibitem [{\citenamefont {Martins}(2019)}]{Martins:2018dqg}%
  \BibitemOpen
  \bibfield  {author} {\bibinfo {author} {\bibfnamefont {C.~J. A.~P.}\
  \bibnamefont {Martins}},\ }\bibfield  {title} {\enquote {\bibinfo {title}
  {{Scaling properties of cosmological axion strings}},}\ }\href {\doibase
  10.1016/j.physletb.2018.11.031} {\bibfield  {journal} {\bibinfo  {journal}
  {Phys. Lett. B}\ }\textbf {\bibinfo {volume} {788}},\ \bibinfo {pages}
  {147--151} (\bibinfo {year} {2019})},\ \Eprint
  {http://arxiv.org/abs/1811.12678} {arXiv:1811.12678 [astro-ph.CO]}
  \BibitemShut {NoStop}%
\bibitem [{\citenamefont {Hindmarsh}\ \emph {et~al.}(2021)\citenamefont
  {Hindmarsh}, \citenamefont {Lizarraga}, \citenamefont {Lopez-Eiguren},\ and\
  \citenamefont {Urrestilla}}]{Hindmarsh:2021vih}%
  \BibitemOpen
  \bibfield  {author} {\bibinfo {author} {\bibfnamefont {Mark}\ \bibnamefont
  {Hindmarsh}}, \bibinfo {author} {\bibfnamefont {Joanes}\ \bibnamefont
  {Lizarraga}}, \bibinfo {author} {\bibfnamefont {Asier}\ \bibnamefont
  {Lopez-Eiguren}}, \ and\ \bibinfo {author} {\bibfnamefont {Jon}\ \bibnamefont
  {Urrestilla}},\ }\bibfield  {title} {\enquote {\bibinfo {title} {{Approach to
  scaling in axion string networks}},}\ }\href {\doibase
  10.1103/PhysRevD.103.103534} {\bibfield  {journal} {\bibinfo  {journal}
  {Phys. Rev. D}\ }\textbf {\bibinfo {volume} {103}},\ \bibinfo {pages}
  {103534} (\bibinfo {year} {2021})},\ \Eprint
  {http://arxiv.org/abs/2102.07723} {arXiv:2102.07723 [astro-ph.CO]}
  \BibitemShut {NoStop}%
\bibitem [{\citenamefont {Chang}\ and\ \citenamefont
  {Cui}(2021)}]{Chang:2021afa}%
  \BibitemOpen
  \bibfield  {author} {\bibinfo {author} {\bibfnamefont {Chia-Feng}\
  \bibnamefont {Chang}}\ and\ \bibinfo {author} {\bibfnamefont {Yanou}\
  \bibnamefont {Cui}},\ }\bibfield  {title} {\enquote {\bibinfo {title}
  {{Gravitational Waves from Global Cosmic Strings and Cosmic Archaeology}},}\
  }\href@noop {} {\  (\bibinfo {year} {2021})},\ \Eprint
  {http://arxiv.org/abs/2106.09746} {arXiv:2106.09746 [hep-ph]} \BibitemShut
  {NoStop}%
\bibitem [{\citenamefont {Hiramatsu}\ \emph {et~al.}(2013)\citenamefont
  {Hiramatsu}, \citenamefont {Kawasaki}, \citenamefont {Saikawa},\ and\
  \citenamefont {Sekiguchi}}]{Hiramatsu:2012sc}%
  \BibitemOpen
  \bibfield  {author} {\bibinfo {author} {\bibfnamefont {Takashi}\ \bibnamefont
  {Hiramatsu}}, \bibinfo {author} {\bibfnamefont {Masahiro}\ \bibnamefont
  {Kawasaki}}, \bibinfo {author} {\bibfnamefont {Ken'ichi}\ \bibnamefont
  {Saikawa}}, \ and\ \bibinfo {author} {\bibfnamefont {Toyokazu}\ \bibnamefont
  {Sekiguchi}},\ }\bibfield  {title} {\enquote {\bibinfo {title} {{Axion
  cosmology with long-lived domain walls}},}\ }\href {\doibase
  10.1088/1475-7516/2013/01/001} {\bibfield  {journal} {\bibinfo  {journal}
  {JCAP}\ }\textbf {\bibinfo {volume} {01}},\ \bibinfo {pages} {001} (\bibinfo
  {year} {2013})},\ \Eprint {http://arxiv.org/abs/1207.3166} {arXiv:1207.3166
  [hep-ph]} \BibitemShut {NoStop}%
\bibitem [{\citenamefont {Aghanim}\ \emph {et~al.}(2020)\citenamefont {Aghanim}
  \emph {et~al.}}]{Aghanim:2018eyx}%
  \BibitemOpen
  \bibfield  {author} {\bibinfo {author} {\bibfnamefont {N.}~\bibnamefont
  {Aghanim}} \emph {et~al.} (\bibinfo {collaboration} {Planck}),\ }\bibfield
  {title} {\enquote {\bibinfo {title} {{Planck 2018 results. VI. Cosmological
  parameters}},}\ }\href {\doibase 10.1051/0004-6361/201833910} {\bibfield
  {journal} {\bibinfo  {journal} {Astron. Astrophys.}\ }\textbf {\bibinfo
  {volume} {641}},\ \bibinfo {pages} {A6} (\bibinfo {year} {2020})},\ \Eprint
  {http://arxiv.org/abs/1807.06209} {arXiv:1807.06209 [astro-ph.CO]}
  \BibitemShut {NoStop}%
\bibitem [{\citenamefont {Figueroa}\ \emph {et~al.}(2020)\citenamefont
  {Figueroa}, \citenamefont {Hindmarsh}, \citenamefont {Lizarraga},\ and\
  \citenamefont {Urrestilla}}]{Figueroa:2020lvo}%
  \BibitemOpen
  \bibfield  {author} {\bibinfo {author} {\bibfnamefont {Daniel~G.}\
  \bibnamefont {Figueroa}}, \bibinfo {author} {\bibfnamefont {Mark}\
  \bibnamefont {Hindmarsh}}, \bibinfo {author} {\bibfnamefont {Joanes}\
  \bibnamefont {Lizarraga}}, \ and\ \bibinfo {author} {\bibfnamefont {Jon}\
  \bibnamefont {Urrestilla}},\ }\bibfield  {title} {\enquote {\bibinfo {title}
  {{Irreducible background of gravitational waves from a cosmic defect network:
  update and comparison of numerical techniques}},}\ }\href {\doibase
  10.1103/PhysRevD.102.103516} {\bibfield  {journal} {\bibinfo  {journal}
  {Phys. Rev. D}\ }\textbf {\bibinfo {volume} {102}},\ \bibinfo {pages}
  {103516} (\bibinfo {year} {2020})},\ \Eprint
  {http://arxiv.org/abs/2007.03337} {arXiv:2007.03337 [astro-ph.CO]}
  \BibitemShut {NoStop}%
\bibitem [{\citenamefont {Gorghetto}\ \emph
  {et~al.}(2021{\natexlab{b}})\citenamefont {Gorghetto}, \citenamefont
  {Hardy},\ and\ \citenamefont {Nicolaescu}}]{Gorghetto:2021fsn}%
  \BibitemOpen
  \bibfield  {author} {\bibinfo {author} {\bibfnamefont {Marco}\ \bibnamefont
  {Gorghetto}}, \bibinfo {author} {\bibfnamefont {Edward}\ \bibnamefont
  {Hardy}}, \ and\ \bibinfo {author} {\bibfnamefont {Horia}\ \bibnamefont
  {Nicolaescu}},\ }\bibfield  {title} {\enquote {\bibinfo {title} {{Observing
  invisible axions with gravitational waves}},}\ }\href {\doibase
  10.1088/1475-7516/2021/06/034} {\bibfield  {journal} {\bibinfo  {journal}
  {JCAP}\ }\textbf {\bibinfo {volume} {06}},\ \bibinfo {pages} {034} (\bibinfo
  {year} {2021}{\natexlab{b}})},\ \Eprint {http://arxiv.org/abs/2101.11007}
  {arXiv:2101.11007 [hep-ph]} \BibitemShut {NoStop}%
\bibitem [{\citenamefont {Braine}\ \emph {et~al.}(2020)\citenamefont {Braine}
  \emph {et~al.}}]{Braine:2019fqb}%
  \BibitemOpen
  \bibfield  {author} {\bibinfo {author} {\bibfnamefont {T.}~\bibnamefont
  {Braine}} \emph {et~al.} (\bibinfo {collaboration} {ADMX}),\ }\bibfield
  {title} {\enquote {\bibinfo {title} {{Extended Search for the Invisible Axion
  with the Axion Dark Matter Experiment}},}\ }\href {\doibase
  10.1103/PhysRevLett.124.101303} {\bibfield  {journal} {\bibinfo  {journal}
  {Phys. Rev. Lett.}\ }\textbf {\bibinfo {volume} {124}},\ \bibinfo {pages}
  {101303} (\bibinfo {year} {2020})},\ \Eprint
  {http://arxiv.org/abs/1910.08638} {arXiv:1910.08638 [hep-ex]} \BibitemShut
  {NoStop}%
\bibitem [{\citenamefont {Backes}\ \emph {et~al.}(2021)\citenamefont {Backes}
  \emph {et~al.}}]{HAYSTAC:2020kwv}%
  \BibitemOpen
  \bibfield  {author} {\bibinfo {author} {\bibfnamefont {K.~M.}\ \bibnamefont
  {Backes}} \emph {et~al.} (\bibinfo {collaboration} {HAYSTAC}),\ }\bibfield
  {title} {\enquote {\bibinfo {title} {{A quantum-enhanced search for dark
  matter axions}},}\ }\href {\doibase 10.1038/s41586-021-03226-7} {\bibfield
  {journal} {\bibinfo  {journal} {Nature}\ }\textbf {\bibinfo {volume} {590}},\
  \bibinfo {pages} {238--242} (\bibinfo {year} {2021})},\ \Eprint
  {http://arxiv.org/abs/2008.01853} {arXiv:2008.01853 [quant-ph]} \BibitemShut
  {NoStop}%
\bibitem [{\citenamefont {Woollett}\ and\ \citenamefont
  {Carosi}(2018)}]{Woollett:2018htk}%
  \BibitemOpen
  \bibfield  {author} {\bibinfo {author} {\bibfnamefont {Nathan}\ \bibnamefont
  {Woollett}}\ and\ \bibinfo {author} {\bibfnamefont {Gianpaolo}\ \bibnamefont
  {Carosi}},\ }\bibfield  {title} {\enquote {\bibinfo {title} {{Photonic Band
  Gap Cavities for a Future ADMX}},}\ }\href {\doibase
  10.1007/978-3-319-92726-8_7} {\bibfield  {journal} {\bibinfo  {journal}
  {Springer Proc. Phys.}\ }\textbf {\bibinfo {volume} {211}},\ \bibinfo {pages}
  {61--65} (\bibinfo {year} {2018})}\BibitemShut {NoStop}%
\bibitem [{\citenamefont {Brun}\ \emph {et~al.}(2019)\citenamefont {Brun} \emph
  {et~al.}}]{MADMAX:2019pub}%
  \BibitemOpen
  \bibfield  {author} {\bibinfo {author} {\bibfnamefont {P.}~\bibnamefont
  {Brun}} \emph {et~al.} (\bibinfo {collaboration} {MADMAX}),\ }\bibfield
  {title} {\enquote {\bibinfo {title} {{A new experimental approach to probe
  QCD axion dark matter in the mass range above 40 $\mu$eV}},}\ }\href
  {\doibase 10.1140/epjc/s10052-019-6683-x} {\bibfield  {journal} {\bibinfo
  {journal} {Eur. Phys. J. C}\ }\textbf {\bibinfo {volume} {79}},\ \bibinfo
  {pages} {186} (\bibinfo {year} {2019})},\ \Eprint
  {http://arxiv.org/abs/1901.07401} {arXiv:1901.07401 [physics.ins-det]}
  \BibitemShut {NoStop}%
\bibitem [{\citenamefont {Beurthey}\ \emph {et~al.}(2020)\citenamefont
  {Beurthey} \emph {et~al.}}]{Beurthey:2020yuq}%
  \BibitemOpen
  \bibfield  {author} {\bibinfo {author} {\bibfnamefont {S.}~\bibnamefont
  {Beurthey}} \emph {et~al.},\ }\bibfield  {title} {\enquote {\bibinfo {title}
  {{MADMAX Status Report}},}\ }\href@noop {} {\  (\bibinfo {year} {2020})},\
  \Eprint {http://arxiv.org/abs/2003.10894} {arXiv:2003.10894
  [physics.ins-det]} \BibitemShut {NoStop}%
\bibitem [{\citenamefont {Lawson}\ \emph {et~al.}(2019)\citenamefont {Lawson},
  \citenamefont {Millar}, \citenamefont {Pancaldi}, \citenamefont
  {Vitagliano},\ and\ \citenamefont {Wilczek}}]{Lawson:2019brd}%
  \BibitemOpen
  \bibfield  {author} {\bibinfo {author} {\bibfnamefont {Matthew}\ \bibnamefont
  {Lawson}}, \bibinfo {author} {\bibfnamefont {Alexander~J.}\ \bibnamefont
  {Millar}}, \bibinfo {author} {\bibfnamefont {Matteo}\ \bibnamefont
  {Pancaldi}}, \bibinfo {author} {\bibfnamefont {Edoardo}\ \bibnamefont
  {Vitagliano}}, \ and\ \bibinfo {author} {\bibfnamefont {Frank}\ \bibnamefont
  {Wilczek}},\ }\bibfield  {title} {\enquote {\bibinfo {title} {{Tunable axion
  plasma haloscopes}},}\ }\href {\doibase 10.1103/PhysRevLett.123.141802}
  {\bibfield  {journal} {\bibinfo  {journal} {Phys. Rev. Lett.}\ }\textbf
  {\bibinfo {volume} {123}},\ \bibinfo {pages} {141802} (\bibinfo {year}
  {2019})},\ \Eprint {http://arxiv.org/abs/1904.11872} {arXiv:1904.11872
  [hep-ph]} \BibitemShut {NoStop}%
\bibitem [{\citenamefont {Pope}(2017)}]{Pope2017}%
  \BibitemOpen
  \bibfield  {author} {\bibinfo {author} {\bibfnamefont {Adrian}\ \bibnamefont
  {Pope}},\ }\href@noop {} {\enquote {\bibinfo {title} {Swfft},}\ }\bibinfo
  {howpublished} {\url{https://xgitlab.cels.anl.gov/hacc/SWFFT}} (\bibinfo
  {year} {2017})\BibitemShut {NoStop}%
\bibitem [{\citenamefont {Wantz}\ and\ \citenamefont
  {Shellard}(2010)}]{Wantz:2009it}%
  \BibitemOpen
  \bibfield  {author} {\bibinfo {author} {\bibfnamefont {Olivier}\ \bibnamefont
  {Wantz}}\ and\ \bibinfo {author} {\bibfnamefont {E.~P.~S.}\ \bibnamefont
  {Shellard}},\ }\bibfield  {title} {\enquote {\bibinfo {title} {{Axion
  Cosmology Revisited}},}\ }\href {\doibase 10.1103/PhysRevD.82.123508}
  {\bibfield  {journal} {\bibinfo  {journal} {Phys. Rev. D}\ }\textbf {\bibinfo
  {volume} {82}},\ \bibinfo {pages} {123508} (\bibinfo {year} {2010})},\
  \Eprint {http://arxiv.org/abs/0910.1066} {arXiv:0910.1066 [astro-ph.CO]}
  \BibitemShut {NoStop}%
\bibitem [{\citenamefont {Borsanyi}\ \emph {et~al.}(2016)\citenamefont
  {Borsanyi} \emph {et~al.}}]{Borsanyi:2016ksw}%
  \BibitemOpen
  \bibfield  {author} {\bibinfo {author} {\bibfnamefont {Sz.}\ \bibnamefont
  {Borsanyi}} \emph {et~al.},\ }\bibfield  {title} {\enquote {\bibinfo {title}
  {{Calculation of the axion mass based on high-temperature lattice quantum
  chromodynamics}},}\ }\href {\doibase 10.1038/nature20115} {\bibfield
  {journal} {\bibinfo  {journal} {Nature}\ }\textbf {\bibinfo {volume} {539}},\
  \bibinfo {pages} {69--71} (\bibinfo {year} {2016})},\ \Eprint
  {http://arxiv.org/abs/1606.07494} {arXiv:1606.07494 [hep-lat]} \BibitemShut
  {NoStop}%
\bibitem [{\citenamefont {Lombardo}\ and\ \citenamefont
  {Trunin}(2020)}]{Lombardo:2020bvn}%
  \BibitemOpen
  \bibfield  {author} {\bibinfo {author} {\bibfnamefont {Maria~Paola}\
  \bibnamefont {Lombardo}}\ and\ \bibinfo {author} {\bibfnamefont {Anton}\
  \bibnamefont {Trunin}},\ }\bibfield  {title} {\enquote {\bibinfo {title}
  {{Topology and axions in QCD}},}\ }\href {\doibase 10.1142/S0217751X20300100}
  {\bibfield  {journal} {\bibinfo  {journal} {Int. J. Mod. Phys. A}\ }\textbf
  {\bibinfo {volume} {35}},\ \bibinfo {pages} {2030010} (\bibinfo {year}
  {2020})},\ \Eprint {http://arxiv.org/abs/2005.06547} {arXiv:2005.06547
  [hep-lat]} \BibitemShut {NoStop}%
\bibitem [{\citenamefont {Tamvakis}\ and\ \citenamefont
  {Wyler}(1982)}]{Tamvakis:1982mw}%
  \BibitemOpen
  \bibfield  {author} {\bibinfo {author} {\bibfnamefont {K.}~\bibnamefont
  {Tamvakis}}\ and\ \bibinfo {author} {\bibfnamefont {D.}~\bibnamefont
  {Wyler}},\ }\bibfield  {title} {\enquote {\bibinfo {title} {{Broken Global
  Symmetries in Supersymmetric Theories}},}\ }\href {\doibase
  10.1016/0370-2693(82)90846-2} {\bibfield  {journal} {\bibinfo  {journal}
  {Phys. Lett. B}\ }\textbf {\bibinfo {volume} {112}},\ \bibinfo {pages}
  {451--454} (\bibinfo {year} {1982})}\BibitemShut {NoStop}%
\bibitem [{\citenamefont {Davis}(1985)}]{Davis:1985pt}%
  \BibitemOpen
  \bibfield  {author} {\bibinfo {author} {\bibfnamefont {Richard~Lynn}\
  \bibnamefont {Davis}},\ }\bibfield  {title} {\enquote {\bibinfo {title}
  {{Goldstone Bosons in String Models of Galaxy Formation}},}\ }\href {\doibase
  10.1103/PhysRevD.32.3172} {\bibfield  {journal} {\bibinfo  {journal} {Phys.
  Rev. D}\ }\textbf {\bibinfo {volume} {32}},\ \bibinfo {pages} {3172}
  (\bibinfo {year} {1985})}\BibitemShut {NoStop}%
\bibitem [{\citenamefont {Vilenkin}\ and\ \citenamefont
  {Vachaspati}(1987)}]{Vilenkin:1986ku}%
  \BibitemOpen
  \bibfield  {author} {\bibinfo {author} {\bibfnamefont {Alexander}\
  \bibnamefont {Vilenkin}}\ and\ \bibinfo {author} {\bibfnamefont {Tanmay}\
  \bibnamefont {Vachaspati}},\ }\bibfield  {title} {\enquote {\bibinfo {title}
  {{Radiation of Goldstone Bosons From Cosmic Strings}},}\ }\href {\doibase
  10.1103/PhysRevD.35.1138} {\bibfield  {journal} {\bibinfo  {journal} {Phys.
  Rev. D}\ }\textbf {\bibinfo {volume} {35}},\ \bibinfo {pages} {1138}
  (\bibinfo {year} {1987})}\BibitemShut {NoStop}%
\end{thebibliography}%
\clearpage

\section*{Methods}

\subsection{Simulation Framework}

We decompose the complex PQ scalar field as $\Phi = (\phi_1 + i \phi_2)/\sqrt{2}$ and assume a radiation-dominated cosmological background. In this notation the axion field is given by $a(x) = f_a \text{arctan2}(\phi_2,\phi_1)$ and the radial mode by $r(x)=\sqrt{\phi_1^2 + \phi_2^2} - f_a$.
The EOM can be derived from the Lagragian in~\eqref{eq:Lagrangian} and expressed in the dimensionless fields $\psi = \phi / f_a $ as
\begin{equation}
\begin{gathered}
\label{eq:EOM}
\psi_1'' + \frac{2}{\eta} \psi_1' - \bar \nabla^2 \psi_1 +   \lambda   \psi _1\left[\eta^2 \left(\psi _1^2+\psi _2^2 - 1\right)+ \frac{T_1^2}{3 f_a^2}\right] = 0, \\
\psi_2'' + \frac{2}{\eta}\psi_2' - \bar \nabla^2 \psi_2 +\lambda \psi _2  \left[\eta^2 \left(\psi _1^2+\psi _2^2 - 1\right)+ \frac{T_1^2}{3 f_a^2}  \right]= 0 \,,
\end{gathered}
\end{equation}
with $T_1$ defined as the temperature when $H(T_1) = f_a$.
Here, primes denote derivatives with respect to $\eta$ while the spatial gradient $\bar \nabla$ is taken with respect to $\bar x = R_1 H_1 x$. 
We chose $\lambda = 1$ without loss of generality and the ratio $(T_1 / f_a)^2$ is given by
\begin{equation}
\left(\frac{T_1}{f_a}\right)^2 \approx 8.4 \times 10^5 \left(\frac{10^{12} \, \mathrm{GeV}}{f_a}\right) \,.
\end{equation}
Note that the PQ breaking scale $f_a$ is degenerate with the choice of physical box size $L$ and dynamical range in $\eta$. This implies that one has to perform only a single simulation, which can be reinterpreted through trivial rescaling for different axion masses.

Using an AMR technique means that some parts of our simulation volume are run at a higher spatial (and temporal) resolution than other parts. 
Our implementation is based on \amrex~\cite{zhang2020amrex}, a software framework for block-structured AMR. 

Our simulation starts out with a uniform grid of $N_{0}=2048^3$ cells, which we will refer to as the {\sl coarse level}.  We generate thermal initial conditions with wavenumber up to 25 in each spatial direction at an initial time $\eta_i = 0.1$. See~\cite{Buschmann:2019icd} for details of how the initial state for $\Phi$ is generated from the thermal correlation functions.  The comoving box length of our simulation volume is $L=120$ with periodic boundary conditions. This implies the simulation contains $(120)^3$ Hubble volumes at $\eta = 1$. Our starting time is $\eta=0.1$. 
Note that the comoving spatial difference $\Delta x_0=L/N_0$ between lattice points is such that our initial state for $\Psi$ is smooth during the initial stages of the PQ phase transition ({\it i.e.}, the structure in $\Psi$ is resolved by multiple grid sites).  

The EOM in~\eqref{eq:EOM} is solved using the strong-stability preserving Runge-Kutta (SSPRK3) method. This method is of third-order and as such one order higher than the often used leapfrog integration scheme. We find that this method provides the best trade-off between numerical stability and computational costs including memory consumption when compared against a second- and fourth-order Runge-Kutta method. At the coarse level, the time step is $\Delta\eta_0=0.02$, satisfying the Courant–Friedrichs–Lewy (CFL) condition at $\Delta \eta_0/ \Delta x_0  \approx 1/3$. The laplacian in the EOM is computed to sufficient accuracy by a second-order finite difference method.

A grid of $N_{0}=2048^3$ cells will not be able to resolve string cores at late times. To maintain resolution we periodically refine a volume around strings, which means decreasing the grid spacing by a factor of 2 in a local volume (see Fig.~\ref{fig:Fig1}).
We refer to the volumes with different resolutions as {\sl levels} $\ell$ with the coarse level being level $\ell=0$. Each level differs from each other not only in spatial resolution, $\Delta x_\ell = \Delta x_0 / 2^\ell$, but also in temporal resolution to locally satisfy the CFL condition, $\Delta\eta_\ell= \Delta\eta_0 / 2^\ell$. 
The higher-resolution lattice on level $\ell$ is determined by fourth-order spatial interpolation of the coarser level $\ell-1$ if no data at that location and level exists. 
Since different grid spacing and time step sizes are used simultaneously, each level is evolved independently and then synchronized appropriately. This is known as the subcycling-in-time approach and requires fourth-order spatial interpolation and second-order temporal interpolation during synchronization. The simulation is insensitive to the exact order of the interpolation used. See the \amrex~documentation~\cite{zhang2020amrex} for more information about the technical details of the AMR approach. 

We add an additional level each time the string width $\Gamma$ drops below four grid sites at the current finest level, $i.e.$ at $\eta\approx 3$, 6, 12, and 24,\footnote{$\log m_r/H\approx2.6$, 3.9, 5.3, 6.7} leading to a total number of 5 levels. A $6^{\rm th}$ level is added at $\eta\approx64$\footnote{$\log m_r/H\approx8.7$} when the string width drops to $3\Delta x_{\ell=4}$. See Supp. Fig.~\ref{fig:resolution} for an illustration of the respective string core resolution at different times in our simulation, compared to the resolution achieved in the static lattice simulation in~\cite{Gorghetto:2020qws}. Note that to match the resolution of the finest level on a uniform grid would require a stunning $65,536^3$ cells.

We use a tagging algorithm to decide on which local volumes to refine, with cells tagged ensured to be within a refined volume. In total we use three different tagging criteria that target $(i)$ string cores, $(ii)$ large gradients in $\psi$, and $(iii)$ short wave-length radiation emitted by strings:

\begin{itemize}
    \item String cores are identified using the procedure described in~\cite{Fleury:2015aca} (Appendix A.2). This involves finding plaquettes that are being pierced by strings. The cell at the low-index corner of a pierced plaquette is tagged.
    \item As strings decay the resulting radiation can produce large gradients in the field. To ensure sufficient resolution we tag cells with $\Delta x_\ell^2\nabla^2\psi_{1,2} > 0.04$. The precise numerical value is of phenomenological origin and has proven to work well in our simulation setup. 
    \item String radiation into radial modes is highly suppressed at late times yet it can cause numerical instabilities if not sufficiently resolved. To avoid a numerical breakdown we tag cells at the coarse level where $\Delta x_0 \eta^3 \nabla^2(\psi_1^2+\psi_2^2) > 4$ is fulfilled.
\end{itemize}

Strings are not stationary  and thus the grid layout has to be adjusted periodically. As this is computationally expensive we re-grid level $\ell$ only every $\Delta \eta_\ell = 0.2 / 2^\ell$. However, we ensure that within this time interval even the fastest moving strings with $v=c$ are always at least a full string width away from any coarse-fine boundary. This is done by leaving a buffer zone of 11 grid sites around each tagged cell that is refined as well. 

The simulation was performed on NERSC's Cori XC40 system using 1024 KNL nodes (Intel Xeon Phi Processor 7250) with, in total, 69,632 physical CPU cores and over 98~TB DDR4 RAM. It ran for about 74 hours ($\sim$5.2 Million CPU hours) in a hybrid OpenMP/MPI mode. 

\subsection{The string length per Hubble $\xi$}

We compute the string length per Hubble $\xi$, defined in the main Article, using the algorithm from~\cite{Fleury:2015aca} that involves counting string-pierced plaquettes; our measured values for $\xi$ are illustrated in Fig.~\ref{fig:xi}.  We then fit the model 
\es{}{
\tilde \xi = {c_{-2} \over \log^2 m_r / H} + {c_{-1} \over \log m_r / H} + c_0 + c_1 \log m_r / H
}
to this data, though this fit is made complicated by the fact that it is difficult to estimate statistical uncertainties from our $\xi$ measurements.  We thus determine these uncertainties in a data-driven way.  Given that we expect the uncertainties to be statistical in nature, and thus determined by the finite simulation volume, we assign uncertainties to each measurement such that the uncertainty at a given $\log_i \equiv \log m_r / H(t_i)$ value is $\sigma_{\xi_i} = \sigma_0 e^{- 3 \log_i / 4} $. Here, the factor $e^{- 3 \log_i / 4}$ is the time-dependence of the square-root of the number of Hubble patches per simulation box, which is a proxy for the square-root of the number of independent string segments in the simulation volume.  We then treat $\sigma_0$ as a nuisance parameter that we profile over during the fit.  In particular, the likelihood is
\es{}{
{\mathcal L}_\xi \left[ {\bm \xi}; {\mathcal M}_\xi, \{ {\bm c}, \sigma_0 \} \right] = \prod_i { \exp \left[ - {(\xi_i - \tilde \xi_i)^2 \over 2 \sigma_{\xi_i}^2 } \right]\over \sqrt{2 \pi \sigma_{\xi_i}^2 }} \,,
}
with $\xi_i$ and $\tilde \xi_i$ denoting the data and model prediction, respectively, at the time labeled by $\log_i$.  Note that we denote the model by ${\mathcal M}_\xi$ with model parameter vector ${\bm c} = \{c_{-2}, c_{-1}, c_0, c_1\}$ in addition to $\sigma_0$.  The uncertainties in Fig.~\ref{fig:xi} arise from the best-fit $\sigma_0$.

\subsection{Construction of the Axion Energy Density Spectrum}
In order to compute the axion energy density spectrum, we consider the screened time-derivative of the axion field, which is defined by
\begin{equation}
    \dot a_\mathrm{scr}(x) = f(x) \dot a(x).
\end{equation}
In this definition, we include a function $f$ that screens out the locations of strings, which appear as discontinuities in the axion field and its derivative. We consider three choices of the screening function:
\begin{align}
    f(x) &= \left[1+r(x)/f_a\right]^2 \label{eq:fid_screening} \\
    f(x) &= 1+r(x)/f_a \label{eq:gio_screening} \\
    f(x) &= 1 \quad\text{(no mask).}\label{eq:no_screening}
\end{align}
In this work our fiducial results use~\eqref{eq:fid_screening} such that $\dot a_\mathrm{scr}(x) = \psi_1(x)\dot{\psi}_2(x) - \dot{\psi}_1(x)\psi_2(x)$. The screening in~\eqref{eq:gio_screening} reproduces that of \cite{Gorghetto:2020qws} while~\eqref{eq:no_screening} corresponds to no string screening. Because $1+r(x)/f_a \approx 1$ at locations far away from string cores, screening  as in~\eqref{eq:fid_screening} and~\eqref{eq:gio_screening} only modify the axion time derivative in the direct vicinity of strings. As shown in Supp. Fig.~\ref{fig:Masking_Fit_Variations}, the results presented in this work are relatively insensitive to the choice of screening function, which can be understood from the fact that we study the emission at spatial scales well beyond the string width.

The axion energy density spectrum within our simulation can then be computed as in \cite{Gorghetto:2020qws} by
\es{eq:spectrum}{
\frac{\partial \rho_a}{\partial k} = \frac{|k|^2}{(2\pi L)^3}\int d\Omega_k |\tilde{\dot{a}}_\text{scr}(k)|^2,
}
where $\tilde{\dot{a}}_\text{scr}(k)$ is the Fourier transform of the field $\dot{a}_\text{scr}$. We compute this energy density spectrum with the HACC SWFFT algorithm \cite{Pope2017} applied to the axion time derivative computed at the coarsest level of spatial resolution. 
After we have computed $d\rho_a/dk$ using the fast Fourier transform (FFT), we then bin our FFT data in 1774 equal-sized bins between $k = 0$ and the maximum $k$, corresponding to $k^\mathrm{max}_\mathrm{com}/2\pi = 1024\sqrt{3} / L$. This binned spectrum is then used in our subsequent analysis.

\subsection{Measuring the string tension}
\label{sec:string_tension}
We compute the effective string tension realized in our simulation following the procedure described in \cite{Gorghetto:2018myk, Gorghetto:2020qws}. We first compute the average energy density within our entire simulation volume using
\begin{equation}
     \rho_\mathrm{tot} =  \langle | \partial \Phi | ^2 + \lambda \left(|\Phi|^2 - \frac{f_a^2}{2}\right)^2 \rangle.
\end{equation}
We then compute the average axion and radial mode energy densities by
\begin{equation}
\begin{gathered}
     \rho_{a} \approx \langle \dot a^2 \rangle \,, \\
     \rho_{r}  \approx \langle \frac{1}{2}\dot r^2 + \frac{1}{2}(\nabla r)^2 + \frac{\lambda}{4} (r^2 + 2 r f_a)^2 \rangle.
\end{gathered}
\end{equation}
In computing $\rho_a$ and $\rho_r$, we mask regions of the simulation volume that are at the highest level of refinement to exclude string contributions. Note that in computing both $\rho_\mathrm{tot}$ and $\rho_r$, we neglect the small contribution of the thermal mass in~\eqref{eq:Lagrangian}. The string energy density is then straightforwardly obtained from
\begin{equation}
\rho_s = \rho_\mathrm{tot} - \rho_a - \rho_r.
\end{equation}

Using the string energy density, we may determine the effective tension by 
\begin{equation}
    \mu_\mathrm{data} = t^2 \rho_s / \xi \,, \label{eq:string_tension}
\end{equation}
with the subscript ``data" denoting the measured value, 
which can be compared to the theoretically expected string tension at large values of $\log m_r / H$:
\begin{equation}
    \mu_\mathrm{th} \approx \pi f_a^2 \log\frac{m_r}{H} \,.
\end{equation}
This comparison is illustrated in Supp.~Fig.~\ref{fig:string_tension} for times between $\log m_r / H= 8$ and $\log m_r / H= 9$. 

Importantly, we only want to compare the leading $\log$ behavior of $\mu_{\rm data}$ and $\mu_{\rm th}$.
Moreover, the addition of a refinement level at $\log m_r / H\approx 8.7$ changes the effective UV cutoff in the numerical calculation, leading to a discontinuity in the measured effective tension.
To analyze the effective tension, we thus adopt a simple logarithmic growth model for the effective tension 
\begin{equation}
\mu = 
\begin{cases} 
      \mu_1 f_a^2 \log m_r / H + \mu_b, &  \log m_r / H \leq 8.7 \\
      \mu_1 f_a^2 \log m_r / H + \mu_a, &  \mathrm{else} \,,
   \end{cases}
\end{equation}
which allows for a different constant offset before ($\mu_b$) and after ($\mu_a$) the addition of the refinement level but enforces uniform logarithmic growth of the string tension. We use a Gaussian likelihood
with data-driven uncertainty on the $\mu_{\rm data}$ values $\sigma_{\mu}$; we treat $\sigma_{\mu}$ as a nuisance parameter in addition to $\mu_{a,b}$.  Profiling over the nuisance parameters we determine $\mu_1 = 3.7 \pm 0.5$, which should be compared to the theoretically expected value $\mu_1 = \pi$. 

\subsection{Instantaneous Emission Analysis}
\label{sec:emission}
Here we describe the method by which we fit a power-law model to the instantaneous emission spectrum. Up to an overall normalization, the instantaneous emission spectrum is given by
\begin{equation}
    F\left(\frac{k}{H}\right) \propto \frac{1}{R^3} \frac{\partial}{\partial t} \left(R^3 \frac{\partial \rho_a}{\partial k} \right).
\end{equation}
In our simulation framework, time evolution is performed in terms of $\eta$ and hence the instantaneous emission $F_i$ at conformal time $\eta_i$ is calculated by numerical finite difference as
\begin{equation}
    F_i\left(\frac{k}{H} \right) \propto \frac{1}{\eta_i^4} \left(\frac{\eta_{i+1}^3 \frac{\partial \rho_{i+1}}{\partial k} - \eta_{i}^3 \frac{\partial \rho_{i}}{\partial k} }{\eta_{i+1}  - \eta_{i}} \right) \,,
\end{equation}
where $\partial \rho_i / \partial k$ is the axion energy density spectrum at $\eta_i$. At each $\eta_i$, we consider a power-law model of the form
\begin{equation}
    f \left(\frac{k}{H}; \{A, q\} \right) = A \left(\frac{k}{H} \right)^{-q}
\end{equation}
and adopt the parametrized form 
\begin{equation}
    \sigma\left(\frac{k}{H}; \{B, p, C\} \right) =  B \left(\frac{k}{H} \right)^{-p} + C
\end{equation}
to describe the combined statistical and systematic uncertainty in the data. We then analyze the data at each $\eta_i$ with the Gaussian likelhood $\mathcal{L}_i$, which is of the form
\begin{equation}
    \mathcal{L}_{i}\left[\mathbf{d}_{i}; \mathcal{M}_{i} \right] = \prod_{j} \frac{1}{\sqrt{2 \pi} \sigma_{j}} \exp\left[-\frac{1}{2}\left(\frac{d_{i, j} - f_{j}}{\sigma_{j}} \right)^2\right]
    \label{eq:power_likelihood}
\end{equation}
where $d_{i, j}$ is the value of the numerically computed instantaneous emission spectrum at the $j^\mathrm{th}$ value of $k/H$ computed at time $\eta_i$. The model predictions for the mean and the error at the $j^\mathrm{th}$ value of $k/H$ are specified by the model parameters $\mathcal{M}_{i} = \{A_i, q_i, B_i, p_i, C_i \}$ for each time $\eta_i$. The values of $k/H$ and associated data that enter the likelihood are restricted to satisfy $k/H > x_\mathrm{IR}$ and $k/H < x_\mathrm{UV}^{-1} m_r / H$. 

In performing the analysis, we only analyze emission spectra which contain at least $10$ bins between $k_\mathrm{IR} \equiv H x_\mathrm{IR}$ and $k_\mathrm{UV} \equiv m_r/x_\mathrm{UV}$. We make the fiducial analysis choices of using the screening function of~\eqref{eq:fid_screening}, $k_{IR} = 50 H$ and $k_{UV} = m_r / 16$, and using a finite difference in time-spacings corresponding to $\Delta \log m_r / H = 0.25$. The impact of varying these fiducial choices, which is marginal, is illustrated in the Supp. Figs.~\ref{fig:UV_Fit_Variations}, \ref{fig:IR_Fit_Variations}, \ref{fig:Log_Fit_Variations}, and \ref{fig:Masking_Fit_Variations}. 

Using the likelihood in~\eqref{eq:power_likelihood}, we determine the maximum likelihood estimate $\hat q_i$ for the emission index at each $\eta_i$. Since the likelihoods are quadratic to very good approximation, we also determine Gaussian uncertainties $\sigma_{q_i}$ on $\hat q_i$ at each $\eta_i$ by $1/\sigma_{q_i}^2 = - \partial^2 /\partial_{q_i}^2 \log \mathcal{L}_i$ evaluated at the likelihood-maximizing model parameters.  After obtaining $\hat q_i$ and $\sigma_{q_i}$ at each $\eta_i$, we join the results to study the possible evolution of $q$. We use a Gaussian likelihood
\begin{equation}
\label{eq:Lq}
    \mathcal{L}_{q} \left[\mathbf{q} ; \mathcal{M}_q, \sigma \right] = \prod_i \frac{\exp \left[-\frac{(\hat q_{i} - \tilde q_{i})^2}{2 (\sigma^2 + \sigma_{q_i}^2)} \right]}{\sqrt{2 \pi (\sigma^2 + \sigma_{q_i}^2})} 
\end{equation}
where $\tilde q_i$ is the model prediction at time $\eta_i$ specified by parameters $\mathcal{M}_q$. We include an additional error term $\sigma$ as a nuisance parameter which is added in quadrature with the data-driven $\sigma_{q_i}$ to address possible systematic effects. In this work, we consider two possibilities for the evolution of $q$, the first that $q$ grows linearly as $q(\log m_r / H) = q_1 (\log m_r / H) +  q_0$ and the second that $q$ is constant such that $q(\log m_r / H) = c_0$. As in our analysis of the individual instantaneous emission spectra, the maximum likelihood estimates and uncertainties of the parameters $\sigma$, $q_0$, and $q_1$ can be determined via standard frequentist techniques.

\subsection{DM abundance calculation}

Here we describe the calculation of the DM abundance from the quantity $n_a^{\rm string}$, which is described in the main Article.
Define $\Lambda \equiv 400$ MeV; then the temperature-dependent axion mass is well characterized by a power-law~\cite{Wantz:2009it}:
\es{}{
m_a^2(T) = {\alpha_a \Lambda^4 \over f_a^2 (T / \Lambda)^n} \,, \qquad  T \gg \Lambda \,,
}
with $\alpha_a$ and $n$ dimensionless constants.  The most recent lattice simulations agree with the dilute instanton gas approximation and support $\alpha_a = (4.6 \pm 0.9)\times 10^{-7}$ for $n \approx 8.16$~\cite{Borsanyi:2016ksw}, which are the values we assume in this work (note that these uncertainties are sub-dominant to those from the axion production from strings from our simulations).  We also approximate the temperature-dependent number of relativistic degrees of freedom as $g_*(T) \approx g_*^0 (T / {\rm MeV})^\gamma$, with $g_*^0 \approx 50.8$ and $\gamma \approx 0.053$, which  
has been shown to match the full result for $g_*(T)$ up to a few percent over the temperature range $800 < T < 1800$ MeV relevant for this calculation~\cite{Lombardo:2020bvn}.  We also assume that the numbers of radiation and entropy degrees of freedom are the same over the temperature range of interest, since the difference between these is also at the level of a few percent~\cite{Borsanyi:2016ksw} and thus a sub-dominant source of uncertainty.

The temperature $T_*$ is defined as the temperature at which $3 H(T_*) = m_a(T_*)$; using $H(T_*) = \pi \sqrt{g_*(T) / 90} T_*^2 / M_{\rm pl}$, with $M_{\rm pl}$ the reduced Planck mass, we may solve explicitly for $T_*$.  We assume that the string network evolves uninterrupted up to $T_*$ but that for $T < T_*$ it quickly evaporates and is not a significance source of axions (but see below). In this approximation axion number density is conserved for $T < T_*$, so that we may write the axion DM abundance today as in~\eqref{eq:omega_a_str}.  Note that $\Omega_a^{\rm str} \propto f_a^{(6+n+\gamma)/(4 + n + \gamma)} \approx f_a^{1.17}$, which is the same scaling as for $\Omega_a^{\rm QCD}$, since for both contributions the $f_a$-scaling has the same origin (see {\it e.g.}~\cite{Buschmann:2019icd}).

While we do not simulate the QCD phase transition in this work, it is important to keep in mind that the string network does evolve non-trivially during the QCD epoch.  As illustrated in the simulations in~\cite{Buschmann:2019icd}, the string network collapses rapidly after $T_*$.  In particular, the string network in~\cite{Buschmann:2019icd} was completely gone at temperatures of order $T \sim T_*/1.5$.  In our approximation where the string network evolves uninterrupted until $T_*$ the string network has energy $\rho_{\rm s} = 4 H^2(T_*) \mu(T_*) \xi_*$ at $T_*$.  Between $T_*$ and $\sim$$1/1.5$$T_*$ all of that energy is transferred to axion radiation.  However, it is likely that the spectrum of radiation during this collapse is shifted to the UV compared to the function $F(k/H)$ from before the mass turns on, since after $T_*$ the axion mass $m_a(T)$ is much larger than Hubble and thus provides an IR cut-off for the radiation spectrum that is further in the UV compared to that for the axion-string network prior to the QCD phase transition.  Since the spectrum is shifted towards the UV, it should produce less axions by number density and thus be less important for the final DM abundance. Still, in order to be conservative we estimate the maximum amount of DM that may be produced by the string network by assuming that at $T_*$ all of the energy density in $\rho_{\rm s}$ is transferred instantaneously to axions with spectrum $F(k/H)$.  This provides a contribution to the axion energy density $\Omega_{\rm str}^{\rm decay} \approx \Omega_a^{\rm str} / 2$, with $\Omega_a^{\rm str}$ being the contribution to the DM abundance from axions produced prior to $T_*$.  We allow for this possibility when determining that the maximum allowed axion mass is $180$ $\mu$eV, but we do not include this contribution when estimating the minimum allowed axion mass of 40 $\mu$eV.

In this work we assume that the radial mass, $m_r$, is of order the decay constant $f_a \sim 10^{10} - 10^{12}$ GeV.
However, one possibility is that $m_r \ll f_a$, as may happen in {\it e.g.} supersymmetric theories where $m_r$ is related to the supersymmetry breaking scale; in this case, $m_r \gtrsim {\rm TeV}$ is possible~\cite{Tamvakis:1982mw}.
If $m_a \sim {\rm TeV}$, then $\log(m_r / H) \sim 50$, which is large enough such that our conclusion that $m_a \in (40, 180)$ $\mu$eV produces the correct DM abundance is still valid in this scenario. 

Note that in these estimates we must perform the fit of the model $\delta \times \sqrt{\xi}$ to the $\langle H / k \rangle^{-1}$ data illustrated in Fig.~\ref{fig:delta}. The $\langle H / k \rangle^{-1}$ data do not have easily estimated uncertainties and so, as we have illustrated multiple times already, we determine these uncertainties in a data-driven way by assigning the uncertainties of all data points a value $\sigma$, which we profile over when determining $\delta$.  The uncertainties in Fig.~\ref{fig:delta} reflect the best-fit value of $\sigma$.

Lastly, the derivation above assumes that number-changing processes are not important in the QCD phase transition since $|a / f_a| \lesssim 1$.  Note that the formula in the main Article for $\langle (a / f_a)^2 \rangle$ for the string-induced axion radiation  arises from the relation \mbox{$\langle (a / f_a)^2 \rangle = (1/f_a^2) \int dk\,  d \rho_a / dk (1/k^2) $}, with \mbox{$d \rho_a / dk = \int^t dt' {\Gamma' / H'} (R' / R)^3 F( k R / R' H')$} and primes denoting quantities evaluated at $t'$.

\subsection{Semi-analytic analysis of string evolution}

In the main Article we pointed to an argument related to the logarithmically-increasing string tension for why $\xi$ may be expected to increase logarithmically in time as well.  Here, we expand upon that argument as well as give an argument for why $q = 1$ may be expected for the spectrum.  As the string network evolves in the scaling regime axions are produced at a rate $\Gamma_a \approx 2 H \rho_s$, where $\rho_s = \xi \mu / t^2$ is the energy density in strings, and $\mu \approx \mu_0 \log(m_r / H)$ is the string tension, to leading order in large $\log$.  Recall that \mbox{$\mu_0 = \pi f_a^2$}.
The tension $\mu$ has a logarithmic divergence that is regulated in the IR by the scale of string curvature $\sim$$H^{-1}$ because of energy associated with the axion field configuration, which wraps around the string.
Physically we may imagine that the long strings are composed of a random walk of smaller segments that we refer to as correlation lengths, which may evolve dynamically and straighten on timescales of order $H^{-1}$.  Denote the number density of correlation lengths as $n_c$.  Then, we may relate $\Gamma_a = n_c dE_c / dt$, where $d E_c / dt$ is the power transferred to axions by the straightening correlation lengths.  Previous studies of collapsing closed string loops and straightening string kinks have shown that the loops and kinks lose energy as $dE / dt = -\alpha f_a^2$, with $\alpha \sim \mathcal{O}(1-10)$, regardless of the loop and kink sizes~\cite{Davis:1985pt,Davis:1986xc,Vilenkin:1986ku,Hagmann:1990mj}.  We assume that the correlation lengths radiate as $dE_c /dt = -\alpha f_a^2$ for some $\alpha \sim 1 - 100$. Solving the energy balance equation then leads to a time-dependent correlation length $L_c \approx \alpha / [ \pi H \log (m_r / H) ]$ for large $\log$.  Let us now assume that there are $\sim$$N_{\rm str}$ strings in total per Hubble patch, with each string composed of a random walk of smaller correlation lengths.  This then implies that at large $\log(m_r / H)$ the parameter $\xi$ scales as $\xi \approx N_{\rm str} (\pi / 4 \alpha) \log(m_r / H)$, which reproduces the observed scaling for $N_{\rm str} \sim {\rm few}$, consistent with the simulation data as illustrated in Fig.~\ref{fig:xi}, and $\alpha \sim \mathcal{O}(10)$.

One of the most important results of this Article is the result that $q \approx 1$, to within $\sim$5\%.  In order to further support this result, we show visually that the string distribution is approaching an attractive solution that supports $q = 1$.
String loops can be characterized by the parameter $n_\ell$, which is the number of string loops with size smaller than $\ell$ at a time $t$, as well as by $\xi_\ell$, which is the total length of string loops with size smaller than $\ell$.
In Supp. Fig.~\ref{fig: xi sub H} we illustrate $\xi_\ell/\xi_\infty$ versus the length $\ell$ at various times, with $\xi_\infty = \xi$.
Visually, it is clear that as time progresses, the string loop distribution approaches an attractor solution, whose validity is extending over an increasingly large range of lengths. This sort of attractor solution for the loop distribution was also found for the fat string approximation in Ref.~\cite{Gorghetto:2018myk} but here we are able to show that this also holds in the physical case.
Given the importance of this distribution, we numerically fit a power law model to the data using the same procedure described in Methods Sec.~\ref{sec:emission}. The treatment of uncertainties and definition of the Gaussian likelihood is analogous to that used for the instantaneous emission spectrum, with a power-law model of the form $\xi_\ell = D \ell^m$. We perform the fit at various times $\eta_i$ to obtain corresponding indices $m_i$. The fitting range is $H\ell/\pi\in(8 H/m_r,1/2)$ to ensure we are within the attractor regime. We only include string loop distributions with at least 8 data points within the fitting range and $\log m_r/H \geq 4$.  The results for $m_i$ are then joint using a Gaussian likelihood identical to that for $q$ in \eqref{eq:Lq} assuming $m$ is time-independent. We find $m = 0.97 \pm 0.03$ with the fit illustrated in Supp. Fig.~\ref{fig: xi sub H}.  

Let us now show that $m = 1$ implies $q = 1$. From a $m = 1$ length distribution, we can calculate the number of strings loops with length between $\ell$ and $\ell + d\ell$ to be
\begin{equation} \label{eq: xi const}
\frac{d \xi_\ell}{d \ell} = \ell \frac{d n_\ell}{d \ell} = D \,,
\end{equation}
for some constant $D$.
We can determine the constant $D$ by using
\es{}{
\rho = \int d\ell \frac{d \rho}{d \ell} = \int d\ell \mu \ell \frac{d n_\ell}{d\ell} = D \mu \ell_\text{max} = \frac{\xi_\text{sub H} \mu}{t^2}\,,
}
leading to $D \approx \xi_\text{sub H} /t^3$ with $\xi_\text{sub H}$ representing the total string length in sub-horizon sized string loops.

We are interested in the spectrum of axions emitted by the string network.  
A string loop of length $\ell$ emits axions dominantly at the fundamental frequency $k \sim 1/\ell$.  
Meanwhile, the string loop radiates energy at a rate $\frac{dE}{dt} = - \alpha f_a^2$.  We can now combine all of this knowledge with~\eqref{eq: xi const} to find
\begin{eqnarray}
F[k/H] \propto \alpha f_a^2 \frac{d n_\ell}{dk} = \frac{\alpha c f_a^2}{k}.
\end{eqnarray}
We thus find that given the attractive behavior seen in Supp. Fig.~\ref{fig: xi sub H}, that the instantaneous spectrum of axions emitted by the network should be approaching $q=1$.  
As a side-note, given this understanding of the string loop distribution, we can easily derive the energy density and spectrum of gravity waves emitted by the network using $dE_{GW} / dt = - \alpha_{GW} G \mu^2$~\cite{Gorghetto:2021fsn}.

Finally, we conclude by giving analytic arguments for why Supp. Fig.~\ref{fig: xi sub H} takes the form that it does.  Namely that at larger lengths, $\xi_\ell \propto \ell$, and at smaller lengths $\xi_\ell \propto \ell^2$.  At small lengths, the string loops shrink due to the emission of axions giving $\ell(t) = \ell_0 - \alpha f_a^2 (t - t_0)/\mu$ with $\ell_0$ being the initial loop size at a time $t_0$.  If string loops are formed at a constant rate with a fixed length $\ell_0$, then $dn_\ell/d\ell \propto dn_{\ell_0}/dt = $ constant.  Multiplying by $\ell$ and integrating, one finds that at small lengths $\xi_\ell \propto \ell^2$, in rough agreement with Supp. Fig.~\ref{fig: xi sub H}.

Larger string loops shrink by intersecting the long, relatively straight, and infinite strings that carry most of the string length.  The two strings will intersect at a rate $\Gamma_{\rm int}$ given roughly by the average string speed over the average distance between strings.  Upon intersecting the infinite string, the string loop loses a random amount of its string length.  If the locations of the intersections are random, the probability distribution for the final length of the string loop, $\ell$, is proportional to its length.  Thus an initial string loop of length $\ell_0$ has $dP/d\ell = 2 \ell/\ell_0^2$.  Putting this intuition into equation form, we find
\begin{equation}
    \frac{dn_\ell}{dt d\ell} = - \Gamma_{\rm int} \frac{dn_\ell}{d\ell} + \int_{\ell}^\infty d\ell_0 \frac{dP}{d\ell} \frac{dn_{\ell_0}}{d\ell_0} \Gamma_{\rm int} \,.
\end{equation}
The first term on the right hand side gives the loss of loops due to intersections while the second term gives their production from larger loops of size $\ell_0$.  Solving for the equilibrium distribution, we find that $dn_\ell/d\ell \propto 1/\ell$.  As before, multiplying by $\ell$ and integrating, one finds that $\xi_\ell \propto \ell$ giving $q=1$.  Meanwhile, the infinite strings are mostly straight except for some highly curved regions that they obtain from intersections with smaller string loops.  As a result, it is natural to expect that the infinite strings radiate axions with the same $q=1$ frequency spectrum as the string loops.

\section*{Data availability}
All data products of this work may be made available by the corresponding authors upon request. Supplementary animations are available at \url{https://bit.ly/amr_axion}.

\section*{Code availability}
The \texttt{AMReX} code framework used in this work is publicly available at \url{https://amrex-codes.github.io/}. Additional code may be made available upon request.

\begin{acknowledgments}
{\it 
We thank  Marco Gorghetto, David Marsh, Javier Redondo, Alejandro Vaquero, Ofri Telem, and Giovanni Villadoro for fruitful discussions.
M.B. was supported by the DOE under Award Number
DESC0007968.  J.F. and B.R.S. were  supported  in  part  by the  DOE  Early  Career  Grant  DESC0019225. This research used resources of the National Energy Research Scientific Computing Center (NERSC), a U.S. Department of Energy Office of Science User Facility located at Lawrence Berkeley National Laboratory, and the Lawrencium computational cluster  provided by the IT Division at the Lawrence Berkeley National Laboratory, both  operated under Contract No. DE-AC02-05CH11231. This research was supported by the Exascale Computing Project (17-SC-20-SC), a collaborative effort of the U.S. Department of Energy Office of Science and the National Nuclear Security Administration.}

\end{acknowledgments}

\clearpage

\onecolumngrid
\begin{center}
  \textbf{\large Supplementary Figures and Tables for Dark Matter from Axion Strings with Adaptive Mesh Refinement}\\[.2cm]
  \vspace{0.05in}
  {Malte Buschmann, \ Joshua W. Foster, \ Anson Hook, \ Adam Peterson, \ Don E. Willcox, \ Weiqun Zhang, \ and Benjamin R. Safdi}
\end{center}

\onecolumngrid
\setcounter{equation}{0}
\setcounter{figure}{0}
\setcounter{table}{0}
\setcounter{section}{0}
\makeatletter
\renewcommand{\theequation}{S\arabic{equation}}
\renewcommand{\thefigure}{S\arabic{figure}}
\renewcommand{\theHfigure}{S\arabic{figure}}
\renewcommand{\thetable}{S\arabic{table}}

\begin{figure}[htb]\centering{
\includegraphics[width =.5 \textwidth]{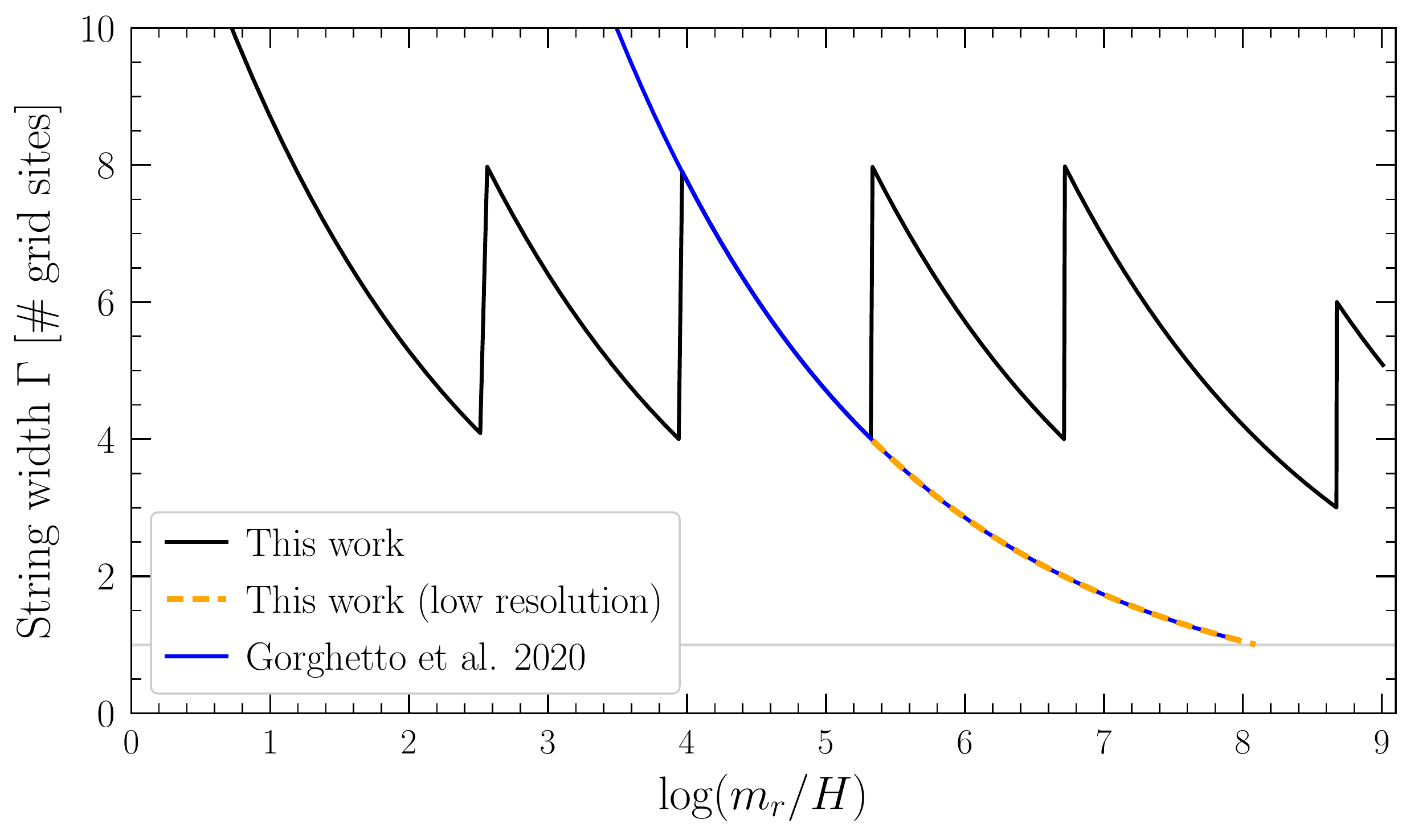}}
\caption{String width $\Gamma$ in units of number of grid sites as a function of simulation time parameterized by $\log m_r/H$. 
Each sudden increase in string width is due to the transformation $\Delta x\rightarrow \Delta x/2$ when an extra level is added. The first four refinement levels are added when $\Gamma/\Delta x=4$ leading to a sudden increase to $\Gamma/\Delta x=8$ each time. 
To test the effect of limited resolution we performed a low-resolution simulation {\sl (dashed red)} that is identical to the main result {\sl (solid black)} up to $\log m_r/H \approx 5.3$ but does not add any extra refinement levels afterwards.
We compare this to the approximate resolution of simulations on a static lattice {\sl (blue)} presented in~\cite{Gorghetto:2020qws}. Simulations on a static lattice over-resolve string cores at early times while under-resolving them at late times. The gray horizontal line corresponds to the often used resolution criteria $m_r\Delta x\lesssim 1$.}
\label{fig:resolution}
\end{figure}

\begin{figure}[htb]
\centering{
\includegraphics[width =.6 \textwidth]{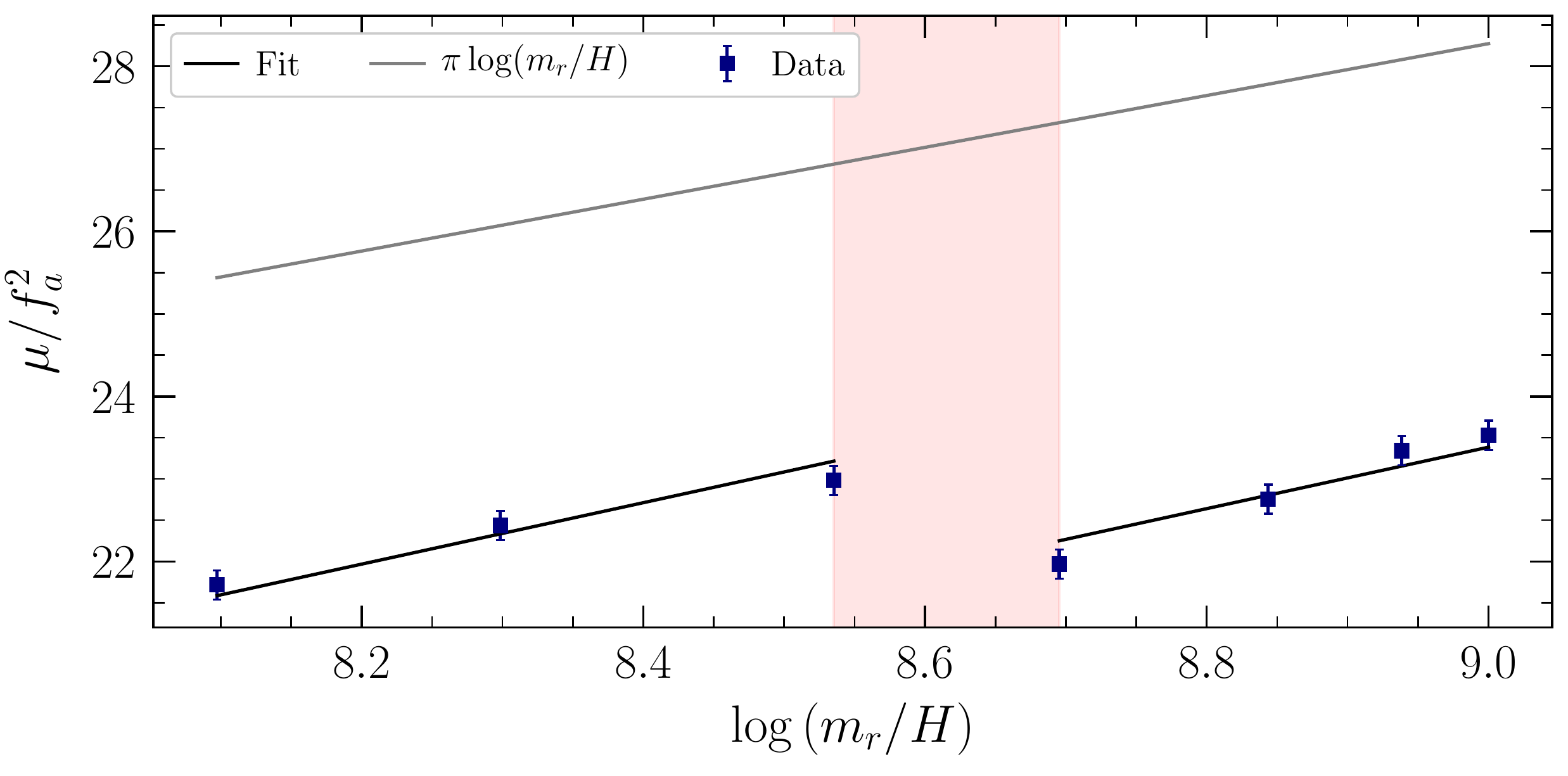}}
\caption{The string tension realized in our simulation as a function of time (see Methods Sec.~\ref{sec:string_tension} for details of how this is computed). The theoretical string tension as a function of $\log m_r / H$ is shown in grey, to leading order in large $\log$, while the string tensions measured in our simulation with data driven errors are shown in dark blue.  Only the leading $\log$ growth of the data is expected to match the theoretical expectation; we find consistency between the linear growth in $\log m_r / H$ of the theoretical and measured string tensions.  Note that an additional refinement level is added during the red band, at $\log m_r / H\approx 8.7$, leading to a change in the overall offset of the $\mu$ data. 
}
\label{fig:string_tension}
\end{figure}

\begin{figure}[htb]
\centering{
\includegraphics[width =.99 \textwidth]{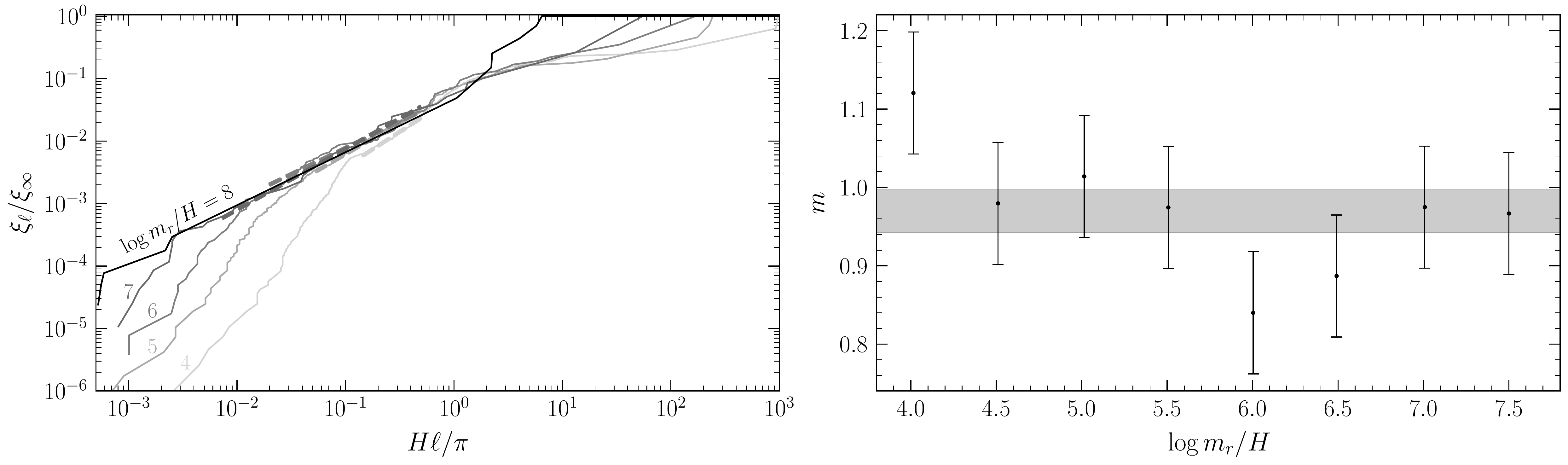}}
\caption{(Left) The total string length in string loops smaller than $\ell$, $\xi_\ell$, versus $\ell$ for various values of $\log m_r/H$.  It is clear that as time progresses the string loop distribution is approaching an attractor solution. We perform a power-law fit of the form $\xi_\ell = D \ell^m$ to the string loop distribution within the attractor regime (dashed lines). (Right) Distribution of the index $m$. The result is joint assuming no time dependence finding $m = 0.97 \pm 0.03$ (gray band).  As argued in the text, this attractor solution leads to an axion spectrum with $q = 1$.}
\label{fig: xi sub H}
\end{figure}

\begin{figure}[htb]\centering{
\includegraphics[width =.99 \textwidth]{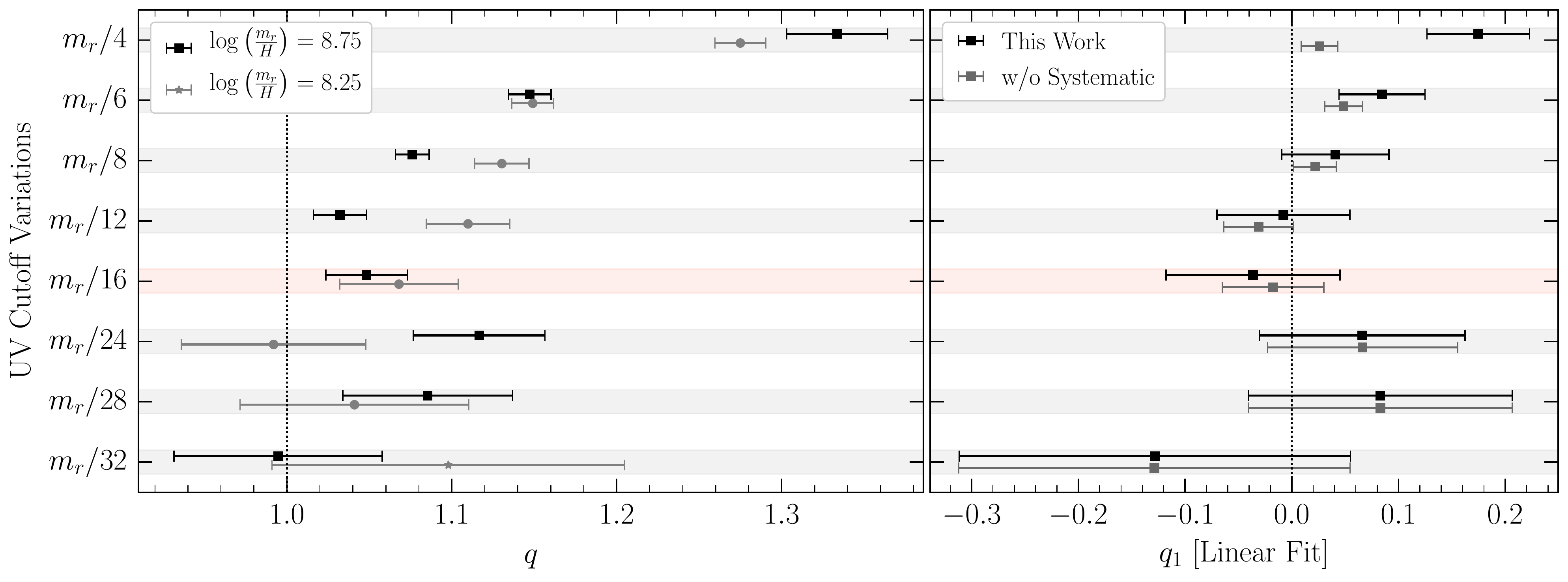}}
\caption{(\textit{Left}) A comparison of the best-fit values for the index $q$ of string emission at selected times for several choices of the UV cutoff of the fit range. (\textit{Right}) Same as {\sl left} but for $q_1$. All index evolution results for these variations are presented in detail in Tab.~\ref{tab:uv_variations}.  Note that the ``w/o Systematic" data points do not include the systematic nuisance parameter $\sigma$ as given in the likelihood in~\eqref{eq:Lq}.}
\label{fig:UV_Fit_Variations}
\end{figure}

\begin{figure}[htb]\centering{
\includegraphics[width =.99 \textwidth]{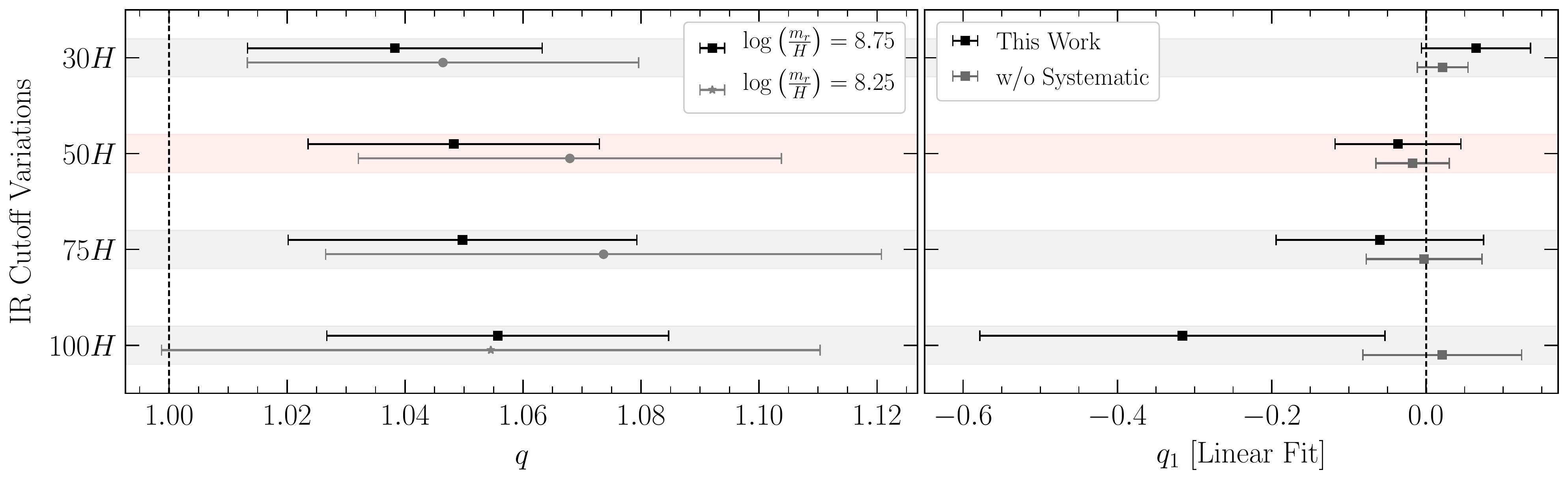}}
\caption{As in Fig.~\ref{fig:UV_Fit_Variations}, but comparing choices of the IR cutoff. All index evolution results for these variations are presented in detail in Tab.~\ref{tab:ir_variations}.}\label{fig:IR_Fit_Variations}
\end{figure}

\begin{figure}[htb]\centering{
\includegraphics[width =.99 \textwidth]{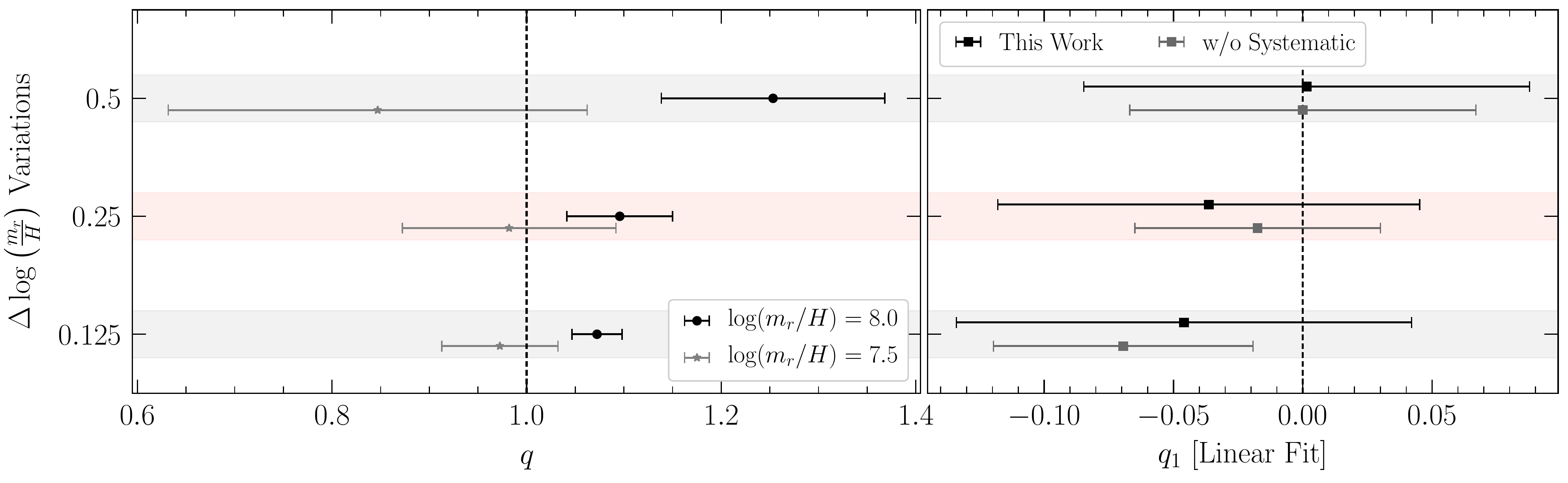}}
\caption{As in Fig.~\ref{fig:UV_Fit_Variations}, but comparing choices of $\Delta \log(m_r / H)$ used in calculating the instantaneous emission spectrum. All index evolution results for these variations are presented in detail in Tab.~\ref{tab:deltaLog_variations}.}\label{fig:Log_Fit_Variations}
\end{figure}

\begin{figure}[htb]\centering{
\includegraphics[width =.99 \textwidth]{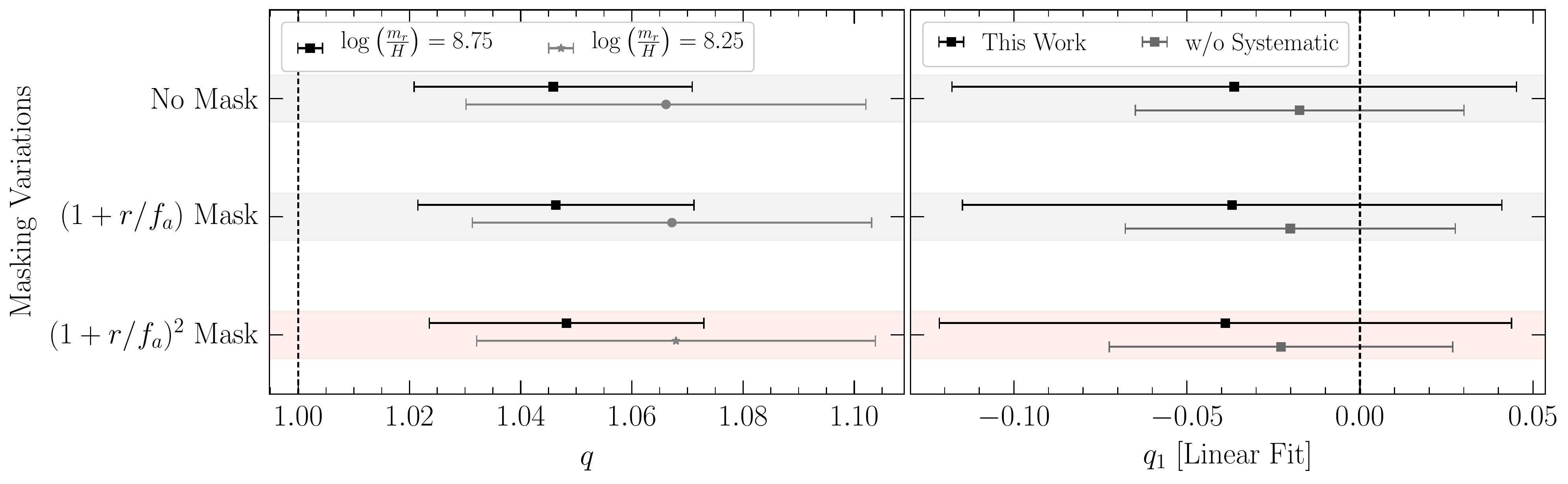}}
\caption{As in Fig.~\ref{fig:UV_Fit_Variations}, but comparing choices of the string masking function. All index evolution results for these variations are presented in detail in Tab.~\ref{tab:masking_variations}.}\label{fig:Masking_Fit_Variations}
\end{figure}

\begin{figure}[htb]\centering{
\includegraphics[width =.99 \textwidth]{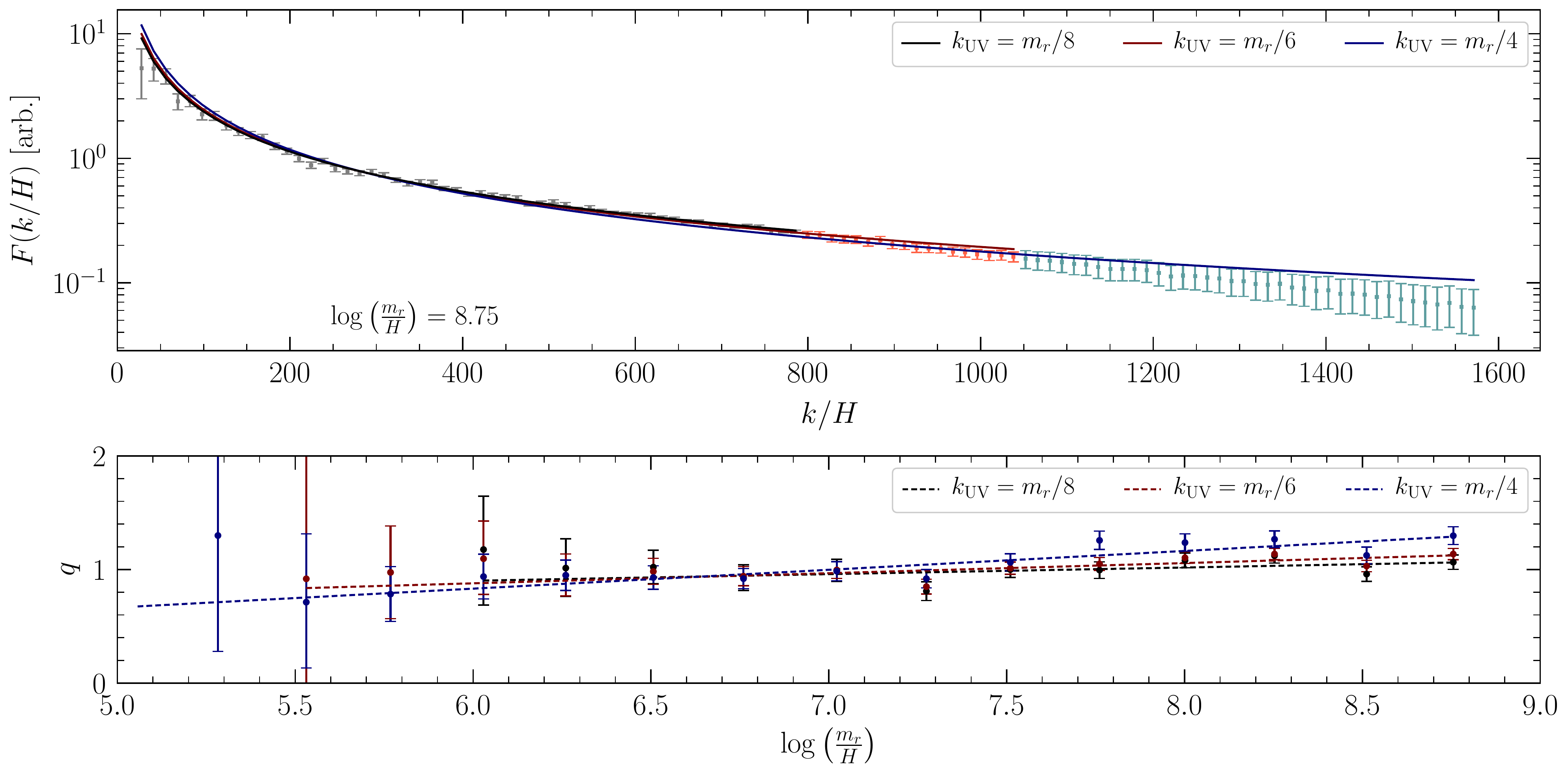}}
\caption{(\textit{Top}) Example fits to the instantaneous emission spectrum at $\log(m_r /H) = 8.75$ for our three largest choices of the UV cutoff for the fitting range. The data indicated in grey corresponds to the range $k \in (30 H, m_r / 8)$, with associated fit in black. In red, which show the fit obtained with the range $k \in (30 H, m_r / 6)$, which includes the grey and additionally light-red data. In dark blue, we show the fit obtained with the range $k \in (30 H, m_r / 4)$, which includes the grey, light-red, and light-blue data. Error bars have been obtained in a data-driven way from the fits using the procedure described in Methods Sec.~\ref{sec:emission}. Visible mismodeling at large $k/H$ biases the fitted power-law towards artificially larger $q$. This can be contrasted with the results shown in Fig.~\ref{fig:SpecIndex}, where a more conservative choice of UV cutoff does not result in apparent mismodeling at an identical time. (\textit{Bottom}) The time evolution of the emission spectrum index for these large choices of UV cutoff for the fitting range. A clear trend of increasing $q$ is obtained for the largest UV cutoff, suggesting that choices of large UV cutoff may result in unphysical growth in the fitted spectral index. Evidence for the linear growth of $q$ in $\log(m_r/H)$ was claimed in \cite{Gorghetto:2020qws} based on analysis performed with the fitting range $k \in (30 H, m_r / 4)$.}\label{fig:Large_UV_Cutoff}
\end{figure}

\begin{figure}[htb]\centering{
\includegraphics[width =.6 \textwidth]{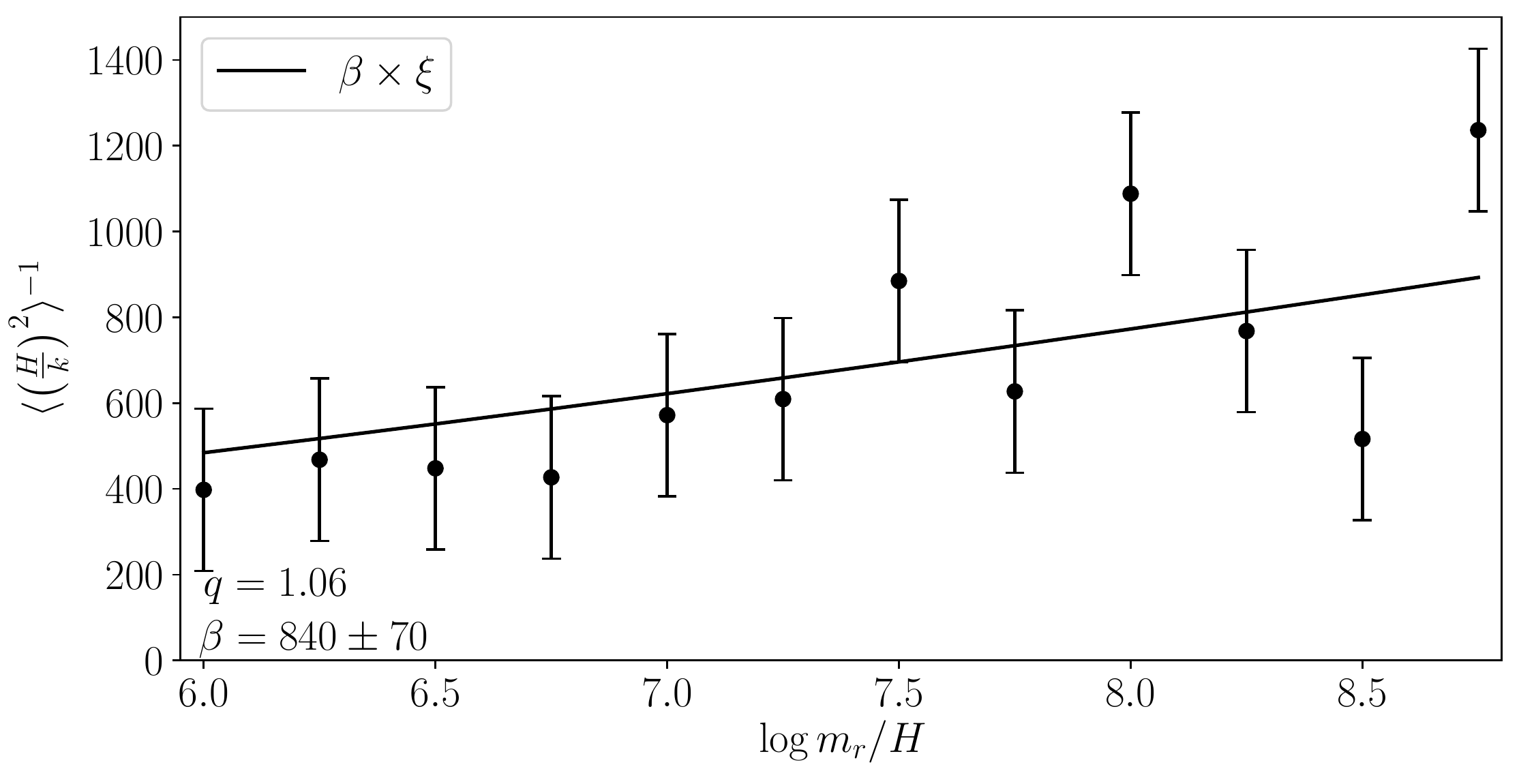}}
\caption{The inverse expectation value $\langle (H / k)^2 \rangle^{-1}$ computed using the axion spectrum $F(k/H)$ by numerically integrating the spectrum to $k/H = x_{\rm max} = 50$ and then analytically integrating the power law distribution $F(x) \propto x^{-q}$ from $x_{\rm max}$ to the UV cut-off at $k/H \sim e^{\log_*}$ for $\log_* \approx 65$ (as in Fig.~\ref{fig:delta}).   Smaller values of $\beta$ correspond to larger axion field values.  Here, we illustrate the result for the maximum allowed $q$ of $1.06$, which leads to the smallest $\beta$ consistent with our simulation results.}
\label{fig:H_over_k2}
\end{figure}

\begin{figure}[htb]\centering{
\includegraphics[width =.95 \textwidth]{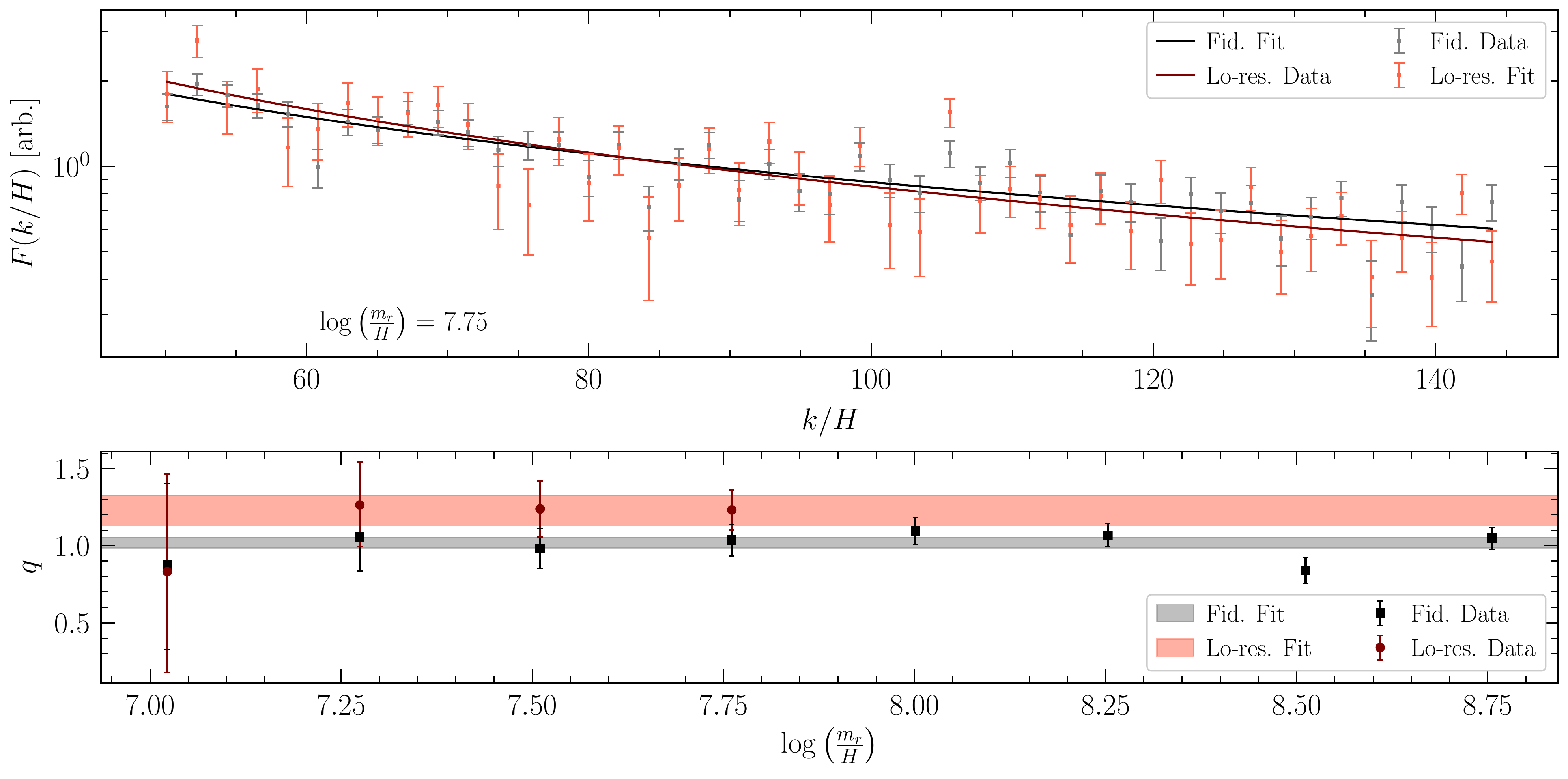}}
\caption{As in Fig.~\ref{fig:SpecIndex}, but comparing fits to the emission spectrum and index evolution for our fiducial simulation output and a lower-resolution simulation. The two simulations are identical until $\log m_r /H \approx 5.3$ when the low-resolution simulation stops adding extra refinement levels. The low-resolution simulation is then run until $\log m_r /H \approx 8$, when we saturate the $m_r \Delta x \lesssim 1$ resolution requirement. (\textit{Top}) A comparison of the emission spectra and fits for the fiducial simulation (data in grey, fit in black) and the lower-resolution simulation (data in light red, fit in maroon) at $\log m_r / H \approx 7.75$, which is the final emission spectrum obtained in our lower-resolution simulation. The lower-resolution simulation prefers a larger power-law index in the fit, and the data-driven errors are somewhat larger than in our fiducial simulation. (\textit{Bottom}) A comparison of the best-fit emission spectrum index as a function of $\log m_r /H$ for the fiducial and lower-resolution simulation. Over the range of $\log m_r / H$ to which we are sensitive in the lower-resolution simulation, the emission spectra realize larger indices, suggesting that resolution loss in a uniform resolution simulation that saturates the resolution criteria may lead to a systematic bias towards index growth.}
\label{fig:LoRes_Comparison}
\end{figure}

\begin{table}[htb]
\ra{1.3}
\centering
\begin{tabularx}{\textwidth}{p{0.2\textwidth}*{5}{P{0.2\textwidth}}}
\hline
  Coefficient & $x_\mathrm{IR} = 30$ & $\mathbf{x_\mathbf{IR} = 50}$ & $x_\mathrm{IR} = 75$ & $x_\mathrm{IR} = 100$\\ \hline
  $q_1$ & $0.07 \pm 0.07$ & $\mathbf{-0.04 \pm 0.08}$ & $-0.06 \pm 0.13$ & $-0.32 \pm 0.26$   \\ \hline
  $q_0$ & $0.41 \pm 0.58$ & $\mathbf{1.36 \pm 0.69}$ & $1.5 \pm 1.12$ & $3.68 \pm 2.21$ \\ \hline
  $q_0^{\rm const.}$ & $0.98 \pm 0.04$ & $\mathbf{1.02 \pm 0.04}$ & $1.0 \pm 0.05$ & $1.02 \pm 0.07$
  \\ \hline
\end{tabularx}
\caption{Results of the fits to the spectral evolution holding all our fiducial analysis choices fixed but for various IR cutoffs $x_\mathrm{IR}$. We provide the fits and uncertainties for the $q_1$ and $q_0$ in the linearly growing index model and the best fit constant for $q_0^{\rm const.}$ in the constant index model. Our fiducial choice of $x_\mathrm{IR} = 50$ is shown in bold.}
\label{tab:ir_variations}
\end{table}

\begin{table}[htb]
\ra{1.3}
\centering
\begin{tabularx}{\textwidth}{p{0.1\textwidth}*{9}{P{0.105\textwidth}}}
\hline
  Coefficient & $x_\mathrm{UV} = 4$ & $x_\mathrm{UV} = 6$ & $x_\mathrm{UV} = 8$ & $x_{UV} = 12$
  & $\mathbf{x_{UV} = 16}$ & $x_\mathrm{UV} = 24$ & $x_\mathrm{UV} = 28$  & $x_\mathrm{UV} = 32$ \\ \hline 
  $q_1$ & $0.17 \pm 0.05$ & $0.08 \pm 0.04$ & $0.04 \pm 0.05$ & $-0.01 \pm 0.06$ & $\mathbf{-0.04 \pm 0.08}$ & $0.08 \pm 0.09$ & $0.08 \pm 0.12$ & $-0.2 \pm 0.2$ \\ \hline
  $q_0$ & $-0.22 \pm 0.37$ & $0.39 \pm 0.32$ & $0.7 \pm 0.41$ & $1.09 \pm 0.51$ & $\mathbf{1.36 \pm 0.69}$ & $0.36 \pm 0.78$ & $0.34 \pm 1.05$ & $2.74 \pm 1.68$ \\ \hline
  $q_0^{\rm const.}$ &  $1.12 \pm 0.05$ & $1.06 \pm 0.03$ & $1.03 \pm 0.03$ & $1.03 \pm 0.03$ & $\mathbf{1.02 \pm 0.04}$ & $1.05 \pm 0.04$ & $1.05 \pm 0.04$ & $1.03 \pm 0.05$  \\ \hline
\end{tabularx}
\caption{As in Tab.~\ref{tab:ir_variations}, but for varying UV cutoff $x_\mathrm{UV}$ with all other parameters fixed to their fiducial values. Our fiducial choice of $x_\mathrm{UV} = 16$ is shown in bold.}
\label{tab:uv_variations}
\end{table}

\begin{table}[htb]
\ra{1.3}
\centering
\begin{tabularx}{\textwidth}{p{0.2\textwidth}*{6}{P{0.2\textwidth}}}
\hline
  Coefficient & $\Delta \log = 0.125$ & $\mathbf{\Delta log = 0.25}$ & $\Delta \log = 0.5$ & $\Delta \log = \log 2$ \\ \hline
  $q_1$ & $0.0 \pm 0.09$ & $\mathbf{-0.04 \pm 0.08}$ & $-0.05 \pm 0.09$ & $-0.1 \pm 0.06$ \\ \hline
  $q_0$ & $1.02 \pm 0.72$ & $\mathbf{1.36 \pm 0.69}$ & $1.4 \pm 0.73$ & $1.84 \pm 0.46$  \\ \hline
  $q_0^{\rm const.}$ & $1.03 \pm 0.04$ & $\mathbf{1.02 \pm 0.04}$ & $1.02 \pm 0.03$ & $1.03 \pm 0.01$ \\ \hline
\end{tabularx}
\caption{As in Tab.~\ref{tab:ir_variations}, but now holding all our fiducial analysis choices fixed, with the exception of the size of the step in $\log(m_r / H)$ used in the finite difference for the calculation of the instantaneous emission spectrum. We vary between $\Delta \log(m_r / H) \in \{0.125, .25, .5, \log(2)\}$, with the $\log(2)$ differences corresponding to a Hubble time spacing. Our fiducial choice of $\Delta \log(m_r / H) = 0.25$ is shown in bold.}
\label{tab:deltaLog_variations}
\end{table}

\begin{table}[htb]
\ra{1.3}
\centering
\begin{tabularx}{\textwidth}{p{0.33\textwidth}*{4}{P{0.225\textwidth}}}
\hline
  Coefficient & \textbf{Eq.~\ref{eq:fid_screening} Mask} &  Eq.~\ref{eq:gio_screening} Mask & Eq.~\ref{eq:no_screening} Mask  \\ \hline
  $q_1$ & $\mathbf{-0.04 \pm 0.08}$ & $-0.04 \pm 0.08$ & $-0.05 \pm 0.08$  \\ \hline
  $q_0$ & $\mathbf{1.36 \pm 0.69}$ & $1.37 \pm 0.65$ & $1.39 \pm 0.7$ \\ \hline
  $q_0^{\rm const.}$ & $\mathbf{1.02 \pm 0.04}$ & $1.02 \pm 0.03$ & $1.02 \pm 0.03$ \\ \hline
\end{tabularx}
\caption{As in Tab.~\ref{tab:ir_variations}, but now holding all our fiducial analysis choices fixed, with the exception of the choice of screening mask. We vary this choice between the screening functions described in~\eqref{eq:fid_screening}, \eqref{eq:gio_screening}, and \eqref{eq:no_screening}. Our fiducial choice of screening in the form of~\eqref{eq:fid_screening} is shown in bold.}
\label{tab:masking_variations}
\end{table}

\clearpage

\end{document}